\documentclass[aps, prl, twocolumn, superscriptaddress]{revtex4-1}
\usepackage{amssymb, amsmath, amsthm}
\usepackage{color}
\usepackage{graphicx}
\usepackage{hyperref}
\usepackage{comment}

\newcommand{\mytitle}{Strong correlations in lossy one-dimensional quantum gases:\\from the quantum Zeno effect to the generalized Gibbs ensemble}

\begin{document}
 
\title{\mytitle}      

\author{Davide Rossini}
\affiliation{Dipartimento  di  Fisica  dell'Universit\`a  di  Pisa  and  INFN,  Largo  Pontecorvo  3,  56127  Pisa,  Italy}
\affiliation{Universit\'e Paris-Saclay, CNRS, LPTMS, 91405 Orsay, France} 

\author{Alexis Ghermaoui}
\affiliation{Laboratoire  Kastler  Brossel,  Coll\`ege  de  France,  CNRS, ENS-PSL  Research  University,  Sorbonne  Universit\'e, 11  Place  Marcelin  Berthelot,  75005  Paris,  France}

\author{Manel Bosch Aguilera}
\altaffiliation{Current address: Department of Physics, University of Basel, Klingelbergstrasse 82, 4056 Basel, Switzerland. }
\affiliation{Laboratoire  Kastler  Brossel,  Coll\`ege  de  France,  CNRS, ENS-PSL  Research  University,  Sorbonne  Universit\'e, 11  Place  Marcelin  Berthelot,  75005  Paris,  France}

\author{R\'emy Vatr\'e}
\affiliation{Laboratoire  Kastler  Brossel,  Coll\`ege  de  France,  CNRS, ENS-PSL  Research  University,  Sorbonne  Universit\'e, 11  Place  Marcelin  Berthelot,  75005  Paris,  France}

\author{Rapha\"el Bouganne}

\author{J\'er\^ome Beugnon}

\author{Fabrice Gerbier}
\affiliation{Laboratoire  Kastler  Brossel,  Coll\`ege  de  France,  CNRS, ENS-PSL  Research  University,  Sorbonne  Universit\'e, 11  Place  Marcelin  Berthelot,  75005  Paris,  France}

\author{Leonardo Mazza}
\email{leonardo.mazza@universite-paris-saclay.fr}
\affiliation{Universit\'e Paris-Saclay, CNRS, LPTMS, 91405 Orsay, France}

\begin{abstract}
We consider strong two-body losses in bosonic gases trapped in one-dimensional optical lattices. 
We exploit the separation of time scales typical of a system in the many-body quantum Zeno regime to establish a connection with the theory of the time-dependent generalized Gibbs ensemble.
Our main result is a simple set of rate equations that capture the simultaneous action of coherent evolution and two-body losses.
This treatment gives an accurate description of the dynamics of a gas prepared in a Mott insulating state and shows that its long-time behaviour deviates significantly from mean-field analyses. 
The possibility of observing our predictions in an experiment with $^{174}$Yb in a metastable state is also discussed.
\end{abstract}

\maketitle

\paragraph{Introduction ---}
Dissipation, noise and losses are ubiquitous in experiments with quantum systems. Although they are typically associated with decoherence~\cite{Zurek_2003}, 
they can also induce interesting  phenomena. 
An iconic example is the quantum Zeno effect, according to which the lifetime of an unstable quantum system can dramatically
increase if it is repeatedly (or even continuously) observed~\cite{Misra_1977, Itano_1990, Facchi_2002}. The same effect also arises for a quantum system dissipatively coupled to an external environment, since this situation can always be interpreted as a generalized (unread) measurement~\cite{Beige_2000,  Almut_2000, Kempe_2001}.  

While earlier studies 
focused on simple quantum systems, there is increasing interest and progress in out-of-equilibrium many-body quantum physics. This field is still in its infancy, but 
several flexible platforms are now available for experimental studies, 
\textit{e.g.}~trapped ions~\cite{Barreiro_2011}, cavity polaritons~\cite{Boulier_2020}, photons in non-linear media~\cite{Carusotto_2020}, and ultra-cold atomic or molecular gases~\cite{Soeding_1999, Tolra_2004, Haller_2011, Schmidutz_2014, Ott_2016, Rauer_2016, Tomita_2017,  Bouganne_2019, Bouchoule_2020}. A major goal is not only to understand quantitatively the effect of decoherence, but also to harness dissipative phenomena to engineer specific quantum states, or even to enhance quantum coherence and  correlations~\cite{Verstraete_2009, Diehl_2008, Roncaglia_2010, Gong_2017, Schemmer_2018, Dogra_2019, Nakagawa_2020}. 
     
Among all sources of dissipation, $n$-body losses ($n \geq 2$) are particularly interesting because they reduce to a $n$-body hard-core constraint~\cite{Daley_2009, Kantian_2009, FossFeig_2012, Nakagawa_2020}.
This effect was demonstrated experimentally with a bosonic one-dimensional gas of molecules subject to two-body losses ($n=2$)~\cite{Syassen_2008}.
Strong losses lead to an emergent behaviour of the molecules as fermionized (hard-core) bosons~\cite{GarciaRipoll_2009}, evidenced by the counter-intuitive \textit{increase} of the lifetime of the gas when two-body losses become stronger. This pioneering experiment demonstrates a paradigmatic instance of the many-body quantum Zeno effect, where the losses are interpreted as fast and unread measurements. This phenomenon has been probed further in ultracold atomic gases with native~\cite{Tomita_2019} or photoassociative~\cite{Tomita_2017} two-body losses, in multi-component fermionic mixtures~\cite{Zhu_2014, Sponselee_2018}, or bosonic systems with three-body losses~\cite{Mark_2020}. 

In this Letter, we study the dynamics of bosonic gases with two-body losses beyond mean-field. We find evidence for an out-of-equilibrium correlated regime at long times caused by the interplay between coherent dynamics and losses. We identify two main experimental signatures as hallmarks of this regime: (i) the decay of the bosonic population as $t^{-  1/2}$ (instead of $1/t$ for the uncorrelated hard-core boson (HCB) gas~\cite{GarciaRipoll_2009}), and (ii) the emergence of peaks centered around $k= 0$ and $ \pi$ in the momentum distribution. To derive these results, we establish and exploit a connection between the many-body quantum Zeno effect and generalized Gibbs ensembles (GGEs) describing the pseudo-thermalization of an isolated quantum system~\cite{Langen_2015,Essler_2016, Cazalilla_2016, Vidmar_2016, Langen_2016, Lange_2018, Mallaya_2019,Caux_2019, Schemmer_2019}.
This connection allows us to derive physically transparent rate equations which give predictions in excellent agreement with numerical exact simulations.

\paragraph{The problem ---}
We consider a one-dimensional bosonic gas trapped in an optical lattice and subject to on-site two-body losses.
The unitary dynamics is governed by a single-band Bose-Hubbard Hamiltonian $H_0 = -J \sum_j (b_j^\dagger b_{j+1} + \mathrm{H.c.})+ (U/2)\sum_j b^{\dagger2}_j b^2_j$, where $b_j^{(\dagger)}$ are bosonic annihilation (creation) operators satisfying canonical commutation relations, while $J$ is the hopping amplitude and $U$ the (repulsive) real part of the on-site interaction strength. The full dynamics is described by a Lindblad master equation for the density matrix $\rho(t)$~\cite{Syassen_2008, GarciaRipoll_2009}: 
\begin{subequations}
\label{Eq:Meq:1}
\begin{align}
 \frac{\rm d}{\mathrm d t} \rho & = \mathcal L[\rho] =  - \frac{i}{\hbar} [H_0, \rho] + \mathcal{D}[\rho]; \label{Eq:LindbladGeneral} \\
 \mathcal{D}[\rho] & = 
 \sum_j L_j \rho L_j^{\dagger}- \frac{1}{2 }  \left\{ L_j^{\dagger} L_j, \rho \right\}.
 \label{Eq:OriginalL}
\end{align}
\end{subequations}
The first term in the right-hand-side of~\eqref{Eq:LindbladGeneral} describes the unitary evolution, and the second term $\mathcal{D}[\rho]$ the dissipative evolution driven by \textit{jump operators}  $L_j$ for each site $j$. 
The jump operators describing two-body losses are $L_j=\sqrt{\gamma_{\rm 2B}/2} \, b_j^2$, where $-\hbar \gamma_\mathrm{2B} / 2$ is the imaginary part of the interaction strength~\footnote{For $J=0$, the population of doubly-occupied sites decays as $p_2(t) = p_2(0) e^{- \gamma_\mathrm{2B} t}$.}. 
The ratio $\gamma_\mathrm{2B}/U$ is typically fixed by the atomic or molecular properties; in contrast, the ratio $\gamma_{\rm2B}/J$ is tunable by several orders of magnitude. 
 
We consider a system that is initially in an atomic-limit Mott insulator ($J=0$) with one atom per site.
The initial state $\rho_0$ is stable under two-body losses for $J = 0$  (indeed $\mathcal{L}[\rho_0]=0$). 
At $t=0$, the lattice depth is lowered ($J>0$). Atoms can tunnel to neighbouring sites and reach unstable configurations with doubly-occupied sites. 
Our goal is to characterize the dissipative dynamics, focusing on readily measurable observables such as the total number of particles $N(t) = \text{Tr}[ \sum_j b^\dagger_j b_j \, \rho(t)]$.

\paragraph{Many-body quantum Zeno effect ---}
We focus on the quantum-Zeno limit of strong dissipation $\hbar \gamma_\mathrm{2B} \gg J$. Roughly speaking, all Fock states with at least one doubly --or higher-- occupied site decay almost immediately on a time scale $\sim \gamma_\mathrm{2B}^{-1}$. 
This decay thus occurs before any substantial coherent dynamics can take place. The subspace of Fock states with at most one boson per lattice site is quasi-stationary and the long-time dynamics takes place in this space of fermionized HCB~\cite{Facchi_2002}. This kinematic constraint results solely from the strong losses, and already shows that they induce non-trivial correlations.

Using the separation of time-scales $\gamma_\mathrm{2B}^{-1} \ll \hbar /J$, Ref.~\cite{GarciaRipoll_2009} proposes an effective Lindblad master equation $\frac{\rm d}{\mathrm dt} \rho = \mathcal L'[\rho]$ that describes the long-time dynamics in the HCB subspace.  
The effective Hamiltonian is $H' = -J \sum_j ( \beta_j^\dagger \beta_{j+1} + {\rm H.c.})$ and corresponds to a tight-binding model of HCB annihilated by the operators $\beta_j$. The effective jump operators take the form of inelastic nearest-neighbor interactions
$L'_j = \sqrt{\Gamma_{\rm eff}} \beta_j(\beta_{j-1}\!+\!\beta_{j+1})$, with
\begin{equation}
 \Gamma_{\rm eff} = \frac{8}{1+\left(\frac{2U}{\hbar \gamma_\mathrm{2B}}\right)^2} \frac{J^2}{\hbar^2 \gamma_\mathrm{2B}}.
\end{equation}
The effective dissipative dynamics is governed by a novel time-scale $\Gamma_{\rm eff}^{-1} \gg \hbar /J \gg \gamma_{\rm 2B }^{-1}$. These inequalities and the scaling $\Gamma_{\rm eff}^{-1} \propto J^2/\gamma_\mathrm{2B}$ are typical of the quantum Zeno regime. 
 
\begin{figure}[t]
 \includegraphics[width=\columnwidth]{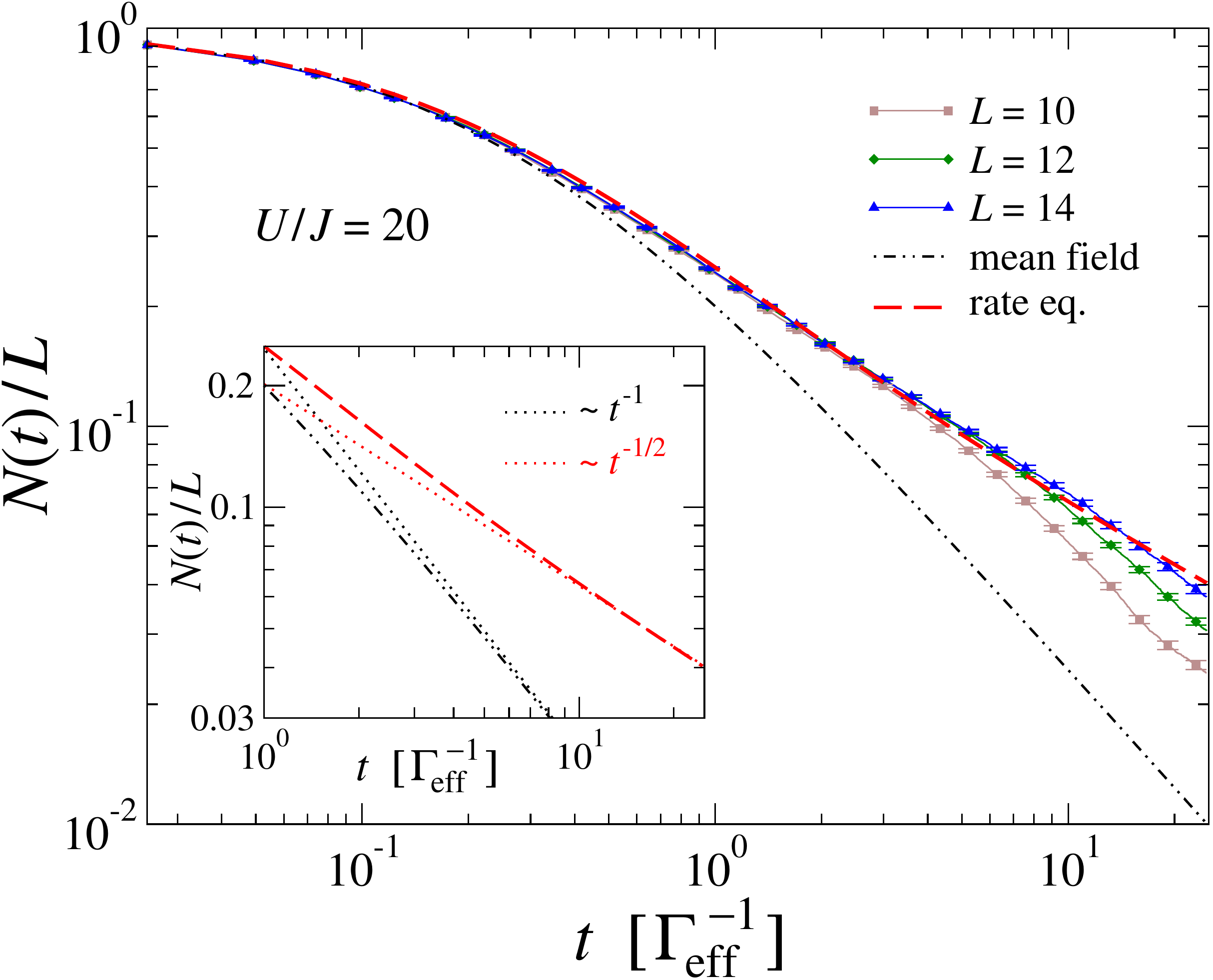}
 \caption{Time-evolution of the number of atoms according to the rate equations~\eqref{Eq:RateEquation} for the initial state $\rho_0$ (dashed red line). We take $U/J=20$ and $U/(\hbar \gamma_{\rm 2B}) = 1.33$ as in $^{174}$Yb.
 Our result is benchmarked with simulations based on quantum trajectories for $L=10$, $12$ and $14$ (each point is averaged over $10^4$ trajectories). 
 The dot-dashed black line represents the mean-field solution $N(t)/L$ in Eq.~\eqref{Eq:MeanField}. The inset highlights the different long-time decay as $t^{-1}$ for the mean-field solution and as $t^{-1/2}$ for the rate equation.} \label{Fig:Numerical1}
\end{figure} 
 
The master equation implies a decay law for the mean atom number $\frac{d \bar{N}}{dt} = -2\sum_j  \langle  L'^{\dagger}_j L'_j \rangle$.
The correlator on the right-hand-side involves inelastic nearest-neighbors interactions $\propto \langle n_j n_{j\pm1}\rangle$ and phase-sensitive density-dependent tunneling $\propto \langle \beta_{j \pm 1}^\dagger n_j \beta_{j\mp 1}\rangle$, 
where 
$n_j=\beta_j^\dagger \beta_j$ is the HCB density. Assuming no correlations between sites, \textit{i.e.}~$\langle  L'^\dagger_j L'_j \rangle \approx 2\langle  n_j  \rangle^2$, Ref.~\cite{GarciaRipoll_2009} derived the mean-field solution,
\begin{equation}
N(t)/L=(1+ 4  \Gamma_{\rm eff} t)^{-1},
\label{Eq:MeanField}
\end{equation}
with $L$ the system length. We note that experimental and numerical data are typically analysed using heuristic modifications of this equation~\cite{Syassen_2008, GarciaRipoll_2009, Sponselee_2018}.
 
In Fig.~\ref{Fig:Numerical1} we compare Eq.~\eqref{Eq:MeanField} with a numerical solution of the HCB model $\frac{\rm d}{\mathrm dt} \rho = \mathcal L'[\rho]$ obtained with state-of-the-art techniques based on quantum trajectories~\cite{Daley_2014} for sizes up to $L=14$. These simulations do not rely on physical approximations 
and serve here as a benchmark. Unsurprisingly, the mean-field solution agrees with the numerics only at short times because the initial state is uncorrelated. Increasingly strong deviations appear at long times, indicating the build-up of correlations that the mean-field model fails to capture.

\paragraph{Rate equations ---}
We now describe our analytical approach to the correlated dissipative dynamics. We interpret the dissipative dynamics as periods of unitary evolution interrupted by quantum jumps where a loss event takes place~\cite{Daley_2014}. Two consecutive loss events are spaced by a time interval $\sim \Gamma_{\rm eff}^{-1}$.  Since the typical time scale of the unitary dynamics of $H'$ is $\hbar/J$, we conclude that according to the inequality $\hbar/J \ll \Gamma_{\rm eff}^{-1}$, the unitary dynamics taking place in between is \textit{long}. 

This dynamics is most easily analyzed after a Jordan-Wigner transformation~\cite{JordanWigner_1928} mapping the HCB to free fermions. Considering periodic boundary conditions, $H'$ then becomes a free fermionic Hamiltonian
$H' = \sum_k \varepsilon(k) c_k^\dagger c_k$, with $k$ the quasi-momentum, $c_k^{(\dagger)}$ canonical fermionic operators, and $\varepsilon(k)=-2 J \cos(k)$. 

The theory of generalized-thermalization in closed quantum systems allows us to describe the state reached after a \textit{long} unitary evolution of $H'$ as a pseudo-thermal state $\sigma$ taking all possible conservation laws into account --  a GGE~\cite{Essler_2016, Cazalilla_2016, Vidmar_2016, Langen_2016}. 
This pseudo-thermal state $\sigma$ is Gaussian in momentum space, thus completely characterised by its correlation matrix $g_{kq}=\mathrm{Tr}[c_k^\dagger c_q \, \sigma]$. The latter is diagonal for a non-interacting and translationally-invariant Fermi gas\,\cite{Sotiriadis_2014},
\begin{equation}
   g_{kq} = \delta_{kq} n_k,
 \label{Eq:LongTime}
\end{equation}
where $\delta_{kq}$ is the Kronecker delta.
We now assume that losses are so rare that the system has enough time in between two loss events to reach a Gaussian generalised-thermal state obeying Eq.~\eqref{Eq:LongTime}. A complete characterization of the dynamics then only requires the knowledge of the occupation number of the different fermionic momenta $n_k(t)=\text{Tr}[c_k^\dagger c_k \rho(t)]$.

We propose to characterise completely the loss dynamics of $N(t)$ by assuming that (i) at every time $t$ the state $\rho(t)$ is Gaussian, and that (ii) it always satisfies momentum factorisation~\eqref{Eq:LongTime}.
Starting from the Lindblad master equation (in the fermionic formulation) and using the aforementioned properties (i) and (ii), we obtain after some algebra the following rate equations~\cite{SupMat}:
\begin{equation}
 \frac{\rm d}{\mathrm d t}n_k(t) = - \frac{4 \Gamma_{\rm eff}}{L}  \sum_q \big[ \sin(k)- \sin(q) \big]^2  n_q(t) \, n_k(t).
 \label{Eq:RateEquation}
\end{equation}
These equations constitute the main result of this article. 
 
\paragraph{Decay of the total number of atoms ---}  
  
The equations~\eqref{Eq:RateEquation} are easily solved numerically. Provided time is properly rescaled in units of $\Gamma_{\rm eff}^{-1}$, we expect that the curves $n_k(t)$ collapse onto a universal function $f_k(x)$ with $x = \Gamma_{\rm eff} t$; similarly, $N(t)$ will collapse onto a function $f(x)$.
The initial state has unit occupation for each momentum, $n_k(0)=1$. In the fermionic representation, it corresponds to a band insulator with the lowest Bloch band entirely filled.
 
We plot in Fig.~\ref{Fig:Numerical1} the density $N(t)/L$ as a function of time for $L=100$ (indistinguishable from the thermodynamic limit, not shown).  
We observe an excellent agreement between the prediction of the rate equation and the numerical simulations for all considered times baring finite size effects. 
We thus conclude that the rate equations~\eqref{Eq:RateEquation}, despite their simplicity, indeed capture the behavior of a complex, interacting and dissipative system. 
Moreover, for a negligible computational cost, they give access to the thermodynamic-limit behaviour.

Unlike the mean field solution, which predicts the scaling $N_0(t) \propto t^{-1}$ at long times, the rate equations~\eqref{Eq:RateEquation} predict that $N(t)$ decays to zero as  $t^{-1/2}$. This result is highlighted in the inset of Fig.~\ref{Fig:Numerical1} and can be analytically proven~\cite{SupMat}. 
This algebraic decay is the hallmark of the correlations that build up after dissipation is enabled. 
   
\begin{figure}[t] 
 \includegraphics[width=\columnwidth]{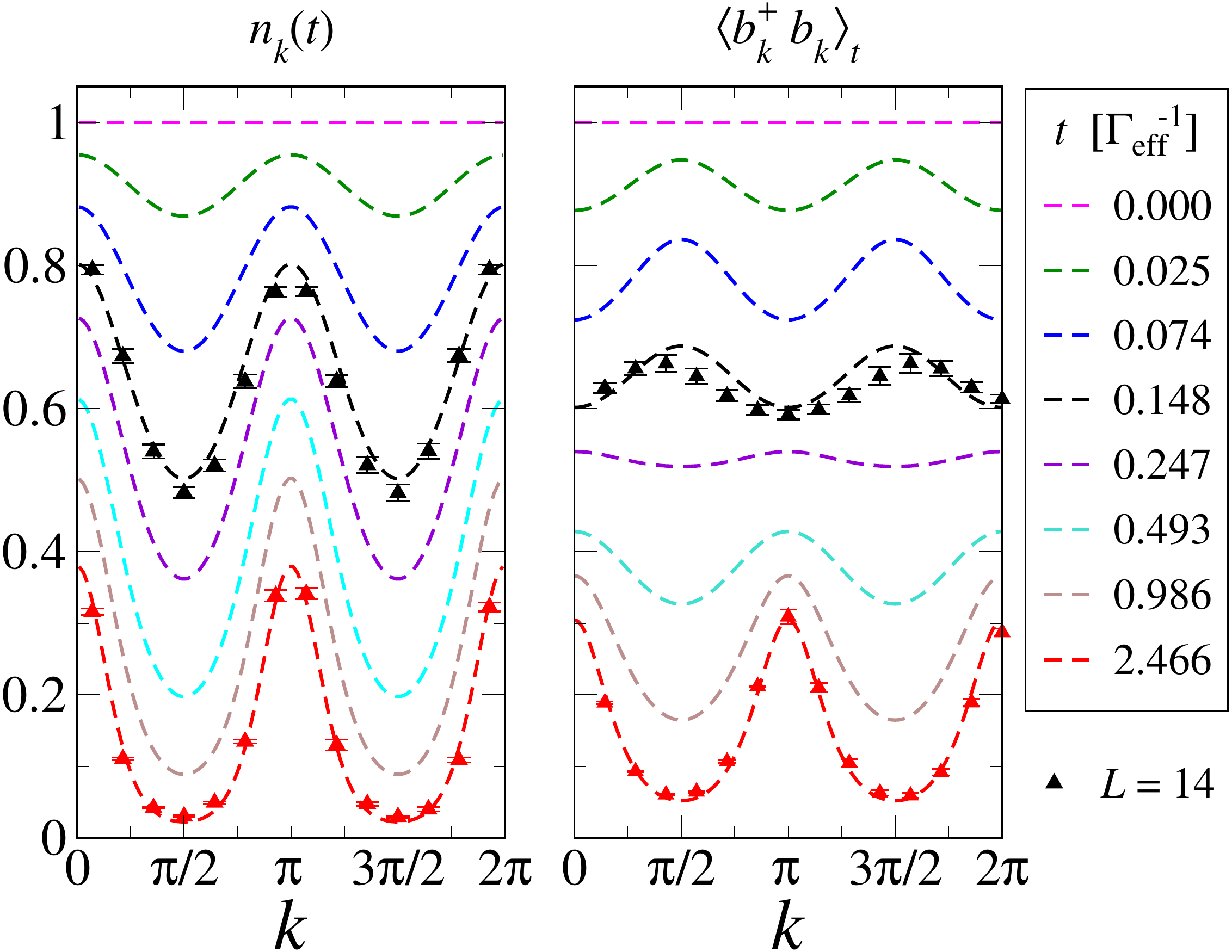} 
 \caption{Fermionic (left) and bosonic (right) quasi-momentum distributions. 
   Dashed lines are the predictions using the rate equation.
   Data from quantum-trajectory simulations for $L=14$ (symbols) are presented for two times.
   }
 \label{Fig:Momentum}
\end{figure}

\paragraph{Momentum distribution function ---}

The rate equations~\eqref{Eq:RateEquation}
provide direct access to the \textit{fermionic} occupation number $n_k(t)$.  
In the Supplementary Material~\cite{SupMat}, we show that the fermionic momentum distribution is well approximated in the long-time limit $t > \Gamma_{\rm eff}^{-1}$ by
\begin{equation}
 n_k(t) \approx \frac{1}{ \left( 8 \pi \Gamma_{\rm eff} t \right)^{1/4}} e^{-\sin^2(k) \, \left( \frac{8 \Gamma_{\rm eff} t}{\pi} \right)^{1/2}}.
 \label{Eq:Fabrice}
\end{equation}
In Fig.~\ref{Fig:Momentum}(left), we plot $ n_k(t)$ for different times, from $t=0$ to $t\sim 2.5 \Gamma_{\rm eff}^{-1}$, and find excellent agreement with the simulations~\cite{SupMat}. Although at initial times the population is uniformly spread among the different momenta, a double-peaked distribution emerges for long times, with maxima at $k = 0, \pi$. The interplay between two-body losses and coherent free-fermion dynamics has thus created a non-equilibrium exotic fermionic gas where the notion of Fermi sea is completely lost.

Standard time-of-flight measurements give instead access to the \textit{bosonic} momentum distribution function $\langle b^\dagger_k b_k\rangle_t$, where $b_k = L^{- \frac 12} \sum_j e^{ikj} b_j$ is a canonical bosonic operator. The link between $\langle b^\dagger_k b_k\rangle_t = \text{Tr}[ b_k^\dagger b_k \, \rho(t)]$ and $n_k(t)$ is known explicitly and we use the approach presented in Ref.~\cite{Gangardt_2006} to compute the distribution shown in Fig.~\ref{Fig:Momentum}(right).
Starting from a flat distribution at $t=0$, the distribution displays two peaks centered around $k = \pm \pi/2$ that persist until the mean density reaches $\bar{n}=0.5$ ($t \leq 0.25 \Gamma_{\rm eff}^{-1}$).
For lower mean densities ($t> 0.25\Gamma_{\rm eff}^{-1}$), peaks appear around $k=0, \pi$, as in the fermionic case.
We compare the results of the rate equations with the exact curves obtained with quantum trajectories for $L=14$. The agreement is excellent at long times and satisfactory at intermediate times $\sim 0.25\Gamma_{\rm eff}^{-1}$. For very short times $t < 0.1 \Gamma_{\rm eff}^{-1}$ the rate equation reproduce poorly the exact data. The numerical calculations show sizeable off-diagonal momentum correlations $\langle c_k^\dagger c_{k'} \rangle$~\cite{SupMat}, implying the failure of the pre-thermalization assumption. 

\paragraph{Time-dependent  GGE ---}

The theory presented so far can be reformulated using the recently-introduced notion of time-dependent GGE (tGGE)~\cite{Lange_2017, Lenarcic_2018, Lange_2018, Mallaya_2019}. 
The interest of this reformulation is conceptual: as originally pointed out in Ref.~\cite{Facchi_2002}, a system in the quantum Zeno regime features quasistationary subspaces and the dynamics constrained therein is generically ruled by a master equation with a strong unitary part and a weak dissipation -- as we are considering here. A tGGE establishes a 
more suitable starting point for the modelisation of other experimental setups~\cite{Zhu_2014, Sponselee_2018, Mark_2020}.

We rewrite the  master equation $\frac{\rm d}{\mathrm dt} \rho = \mathcal L'[\rho]$ as $\frac{\rm d}{\mathrm dt} \rho = \mathcal L_0[\rho] + \mathcal L_1[\rho]$. Here  $\mathcal L_0[\rho]=- \frac{i}{\hbar} [H', \rho]$ describes the dominant unitary dynamics ($\mathcal L_0 \propto J/\hbar$), and $\mathcal L_1$ the weaker dissipative part ($\mathcal L_1\propto\Gamma_{\rm eff} $). To lowest order in $\hbar \Gamma_{\rm eff}/J \ll 1$, the tGGE theory predicts that the system remains at all times in one of the many stationary states of $\mathcal L_0$. The effect of the dissipation $\mathcal L_1$ is then to determine the dynamics within this subspace. 

The tGGE theory relies on a particular \textit{ansatz} for the density matrix. Instead of \textit{all} possible stationary states, the \textit{ansatz} retains only the GGEs for the strong Hamiltonian $H'$,
\begin{equation}
 \rho_{\rm tGGE}(t) = \frac{1}{\mathcal Z(t)} e^{- \sum_k \mu_k(t) c_k^\dagger c_k},
 \label{Eq:tGGE}
\end{equation} 
with time-dependent Lagrange multipliers $\mu_k(t)$ and a generalized partition function $\mathcal Z(t) = \prod_k \big[1+ e^{- \mu_k(t)} \big]$. 
The equations of motion for the $\mu_k(t)$ derived in Ref.~\cite{Lange_2018} describe how $\mathcal L_1$ forces the system to explore different GGE states. 
In our case, we obtain~\cite{SupMat}:
\begin{equation}
 \frac{\rm d}{\mathrm dt} \mu_k(t) = \frac{4 \Gamma_{\rm eff}}{L} \sum_q \big[ \sin(k)-\sin(q) \big]^2 \frac{e^{- \mu_k(t)}+1}{e^{\mu_q(t)}+1} .
 \label{Eq:mu:tGGE}
\end{equation}
The individual occupation numbers for the state~\eqref{Eq:tGGE} obey a Fermi-Dirac law $n_k(t) = (e^{\mu_k(t)}+1)^{-1} $. 
Substituting this expression in Eq.~\eqref{Eq:mu:tGGE}, we recover the rate equations~\eqref{Eq:RateEquation} for $n_k(t)$, thereby establishing the equivalence of the two formulations.

\begin{figure}[t]
\includegraphics[width=\columnwidth]{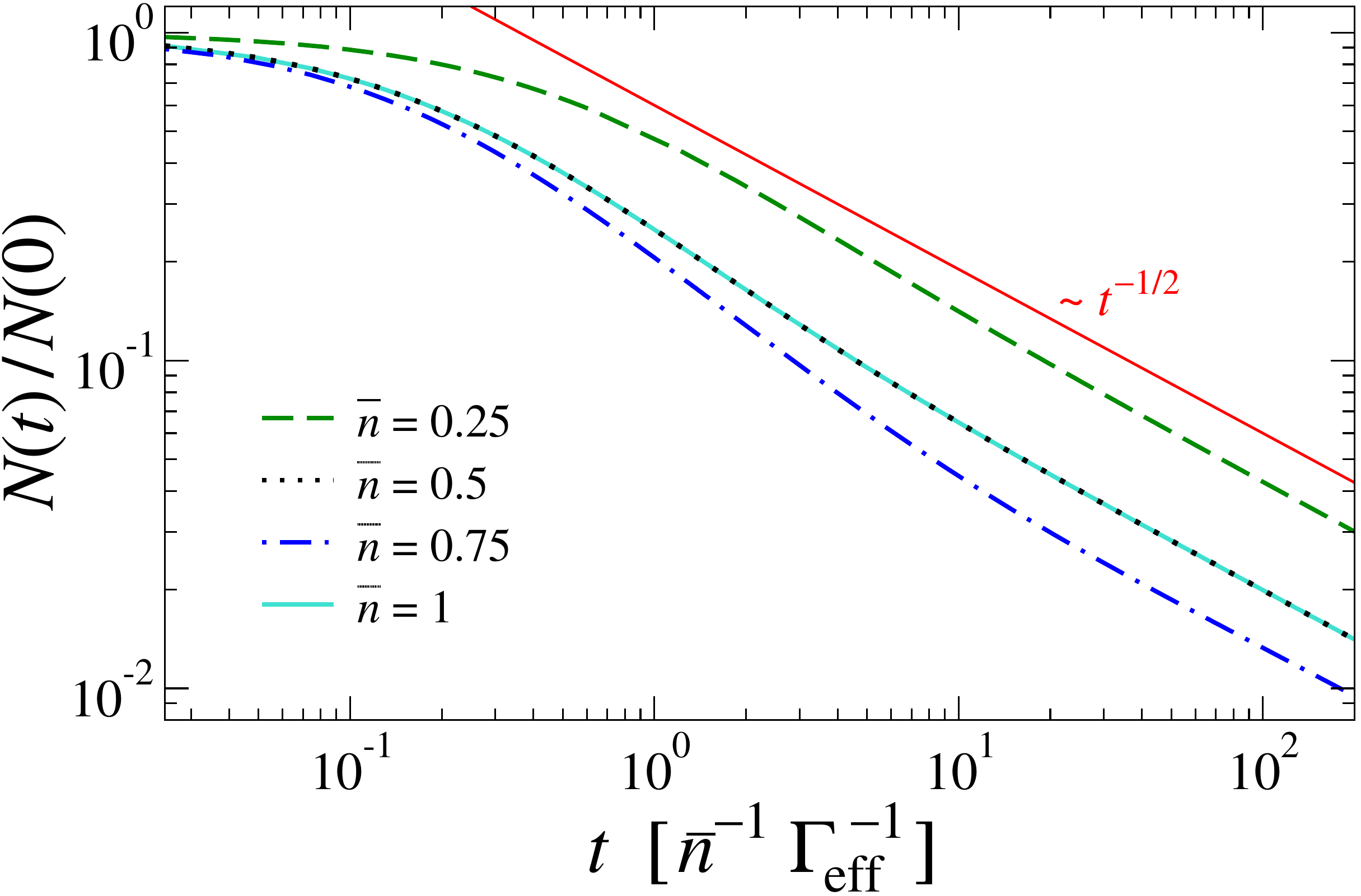}
\caption{Decay of atom number for lattice Tonks-Girardeau gases with density $\bar n \leq 1$. The various curves are calculated according to the rate equations~\eqref{Eq:RateEquation} for different initial conditions. The initial state is taken to be a lattice Tonks-Girardeau gas with $n_k(t=0)$ given by the Fermi-Dirac distribution at zero temperature with mean density $\bar n$.}
\label{Fig:LowDensity}
\end{figure}

\paragraph{Initial state ---} 

The behavior discussed so far is not specific to a Mott insulator initial state with density $\bar n=1$, but is also observed for lower initial fillings. Let us first consider a bosonic gas with equally populated momenta $\langle b_k^\dagger b_k \rangle = \bar n <1$, which maps to $n_k(0) = \bar n$. The rate equations can be solved with a proper rescaling of time $t \to t \bar n$, so that $n_k(t) = \bar n f_k(\bar n \Gamma_{\rm eff} t)$. Thus, for a lower initial density, the loss dynamics simply slows down and the effective decay rate is rescaled by the density.

To model a situation closer to experimental reality,
we now consider an initial state that is the ground state of the Hamiltonian $H'$ with density $\bar n<1$ (a Tonks-Girardeau gas on a lattice~\cite{Girardeau_1960}).   In the fermionic formulation, the initial conditions are determined by Fermi-Dirac statistics $n_k(0) = n_{\rm FD}(k) = (e^{\beta(-2 J  \cos k - \mu)}+1)^{-1}$ with $\beta \to +\infty$. The numerical analysis presented in Fig.~\ref{Fig:LowDensity} shows the results of the rate equation with a rescaling $t \to t \bar n$. We observe that $\bar n=0.5$ and $\bar n=1$ collapse exactly whereas for $\bar n<0.5$ the dynamics slows down.   
On the contrary, for values $0.5<\bar n<1$ the dynamics is slightly faster and non-monotonic in the density. Thus, the decay can be accelerated or decelerated depending on the initial density. In all cases, however, we observe a long-time decay $N(t)\sim t^{-1/2}$. This robust feature of a slower decay thus remains the strongest evidence for the interplay between correlations and losses beyond the mean-field description.

\paragraph{Conclusions and perspectives ---}
We have proposed a novel theoretical approach to the dynamics of a lossy bosonic gas in the many-body quantum Zeno regime.
The quasistationary subspace enables for a theoretical treatment based on generalized thermalisation. 

From an experimental viewpoint, the discussed dynamics can be investigated with any atomic or molecular species featuring strong two-body losses~\cite{Syassen_2008, Zhu_2014, Tomita_2017, Sponselee_2018}, or possibly in other systems as well (for instance, photonic systems with two-photon absorption~\cite{Carusotto_2020}). 
We can estimate the relevant time scales for an optical lattice of $8$ recoil energies ($U/J \sim 20$) loaded with $^{174}$Yb in its metastable excited state (on-site two-body losses have been characterised in Refs.~\cite{Bouganne_2017, Franchi_2017}).
We obtain $U/\hbar = 7,800$ s$^{-1}$, $\gamma_{\rm 2B} = 5,900$ s$^{-1}$ and $J/\hbar= 377$ s$^{-1}$ and as a result $\Gamma_{\rm eff} = 24$ s$^{-1}$. Thus, our predictions require an observation time of $20 \Gamma_{\rm eff}^{-1} \sim 1$s which is within current experimental possibilities. 

Since the conservation of the momentum occupation numbers in-between loss events plays a crucial role, an experimental difficulty is the realization of a truly homogeneous system. Although this has been already achieved experimentally~\cite{Mazurenko2017a}, the vast majority of experiments also include an additional harmonic confinement~\cite{Bloch2008a}. Adapting the rate equation approach to inhomogeneous, harmonically confined systems is an important extension left for future work. Another avenue comes from the tGGE formulation of the dynamics. This establishes a suitable starting point to describe, \textit{e.g.}~fermions with two-body losses~\cite{Zhu_2014, Sponselee_2018} or bosons with three-body losses~\cite{Mark_2020} and to explore two- and three-dimensional systems.
 
\paragraph{Related article ---} While completing this paper, we became aware of a work discussing losses in one-dimensional bosonic gases without lattice~\cite{Bouchoule_2020b}.
   
\paragraph{Acknowledgement ---}
We are grateful to J.~De Nardis for discussions on the bosonic momentum distribution function and to I.~Bouchoule for insightful comments. 
We also acknowledge discussions with A. De Luca, M. Fagotti, R. Fazio, G. La Rocca, L. Rosso, and M. Schir\`o. D.R.~acknowledges hospitality from LPTMS through CNRS funding. 
This work has been partially funded by  LabEx PALM (ANR-10-LABX-0039-PALM).

\newpage 

\clearpage

\onecolumngrid

\begin{center}
\begin{LARGE}
\textbf{Supplementary material}
\end{LARGE}
\end{center}

\setcounter{secnumdepth}{2} 

\section{Derivation of the rate equation}

In this Section we detail the derivation of the rate equations~(5) in the main text. We consider a ring system with periodic boundary conditions throughout.

\subsection{The master equation: fermionization and momentum-space representation}

For reading convenience, we report here the complete master equation for hardcore bosons derived in Ref.~\cite{GarciaRipoll_2009}:
\begin{equation}
  \frac{\rm d}{\mathrm d t} \rho(t) = - \frac{i}{\hbar}
  \big[ H_1 + H_2,\rho(t) \big] + 
  \sum_j \left[ L_j \rho(t) L_j^\dagger - \frac{1}{2} \left\{ L_j^\dagger L_j, \rho(t) \right\} \right];
  \label{Eq:SUPP:MEQ}
\end{equation}
where
\begin{subequations}
  \begin{align}
    H_1 &= - J \sum_j \left( \beta_j^\dagger \beta_{j+1}+ \beta_{j+1}^\dagger \beta_j \right),\\
    H_2 & = -J_{2} \sum_j L_j^\dagger L_j,\\
    L_j &= \sqrt{\Gamma_{\rm eff}} \; \beta_j \big( \beta_{j+1}+ \beta_{j-1} \big).
 \end{align}
\end{subequations}
The $\beta_j^{(\dagger)}$ are hardcore bosons operators obeying the commutation relation $[\beta_j,\beta_j^\dagger]=1-2 n_j$, with $n_j=\beta_j^\dagger\beta_j$. The term $H_2$ is a Hamiltonian correction omitted in the main text (in Sec.~\ref{Sec:No:H2} we comment on the fact that it is irrelevant for our study). The coefficient $J_2$ and the dissipation rate read:
  \begin{equation}
    J_2 = \frac{U}{ \gamma},  \qquad \qquad
    \Gamma_{\rm eff} = \frac{8}{1+\left(\frac{2U}{\hbar \gamma}\right)^2} \frac{J^2}{\hbar^2 \gamma} .
 \end{equation}
In the main text we have called this master equation $\frac{\rm d}{\mathrm d t}\rho = \mathcal L'[\rho]$; the jump operators are called $L'_j$, the Hamiltonian $H_1$ is called $H'$ and $H_2$ is not included because it does not play any role in the dynamics of our interest (see below Sec.~\ref{Sec:No:H2}).

By means of a Jordan-Wigner transformation, 
\begin{equation}
  \left\{ 
  \begin{array}{l}
    c_j \doteqdot e^{i \pi \sum_{m<j} \beta_m^\dagger \beta_m} \, \beta_j; \\
    \beta_j \doteqdot e^{i \pi \sum_{m<j} c^\dagger_m c_m} \, c_j;
  \end{array}
  \right.,
\end{equation}
we map the hardcore-boson problem to a fermionic one and introduce fermionic operators $c_j^{(\dagger)}$ with anticommutation relations $\{  c_i,  c_j \} = 0$  and $ \{ c_i, c_j^\dagger \} =\delta_{i,j}$.

The fermionic representation of the Hamiltonian $H_1$ and of the jump operators $ L_j$ will be useful in the following:
\begin{subequations}
  \begin{align}
    H_1 =&  - J \sum_{j=1}^{L-1} \left( c_j^\dagger  c_{j+1}+  c_{j+1}^\dagger  c_j \right) ; \\
    L_j =&   - \sqrt{\Gamma_{\rm eff}} c_j \left( c_{j+1} - c_{j-1}\right).
  \end{align}
\end{subequations}
The Hamiltonian $H_1$ takes the particularly expressive form of a free-fermion Hamiltonian. The jump operators $L_j$ involve instead nearest-neighbor inelastic loss processes with a sign change with respect to the original bosonic problem. 
The fermionic representation of $H_2$ follows from that of $L_j$ and will not be explicitly specified.

Since our problem is invariant under discrete translations, we can introduce the (quasi-)momentum representation
\begin{equation}
  c_j = \frac{1}{\sqrt{L}} \sum_k e^{i kj} c_k,
\end{equation}
where $j$ labels the site position along the ring and $k$ is the quasi-momentum. The Hamiltonian $H_1$ and the dissipators $L_j$ read:
\begin{subequations}
  \begin{align}
    H_1 =& \sum_k -2 J \cos(k) c_k^\dagger c_k; \\
    L_j =& - \frac{\sqrt{\Gamma_{\rm eff}}}{L} \sum_{k,q} e^{i (k+q) j} \left( e^{ik }- e^{- ik}  \right) c_k c_q = - \frac{\sqrt{\Gamma_{\rm eff}}}{L} \sum_{k,q}  2 i e^{i (k+q) j} \sin(k) \, c_k c_q.
    \label{Eq:Lj:k}
  \end{align} 
  \label{Eq:L:Mom}
\end{subequations}

\textbf{Note:} \textit{We note for the sake of completeness that for a finite-size chain of length $L$, corrective boundary terms arise: 
\begin{subequations}
  \begin{align}
    \delta H_1 =& (-1)^N J \left(  c_L^\dagger  c_{1}+  c_{1}^\dagger  c_L \right); \\
    L_j =& \left\{ 
    \begin{array}{ll}
      - \sqrt{\Gamma_{\rm eff}} \;  c_j \left[ c_{j+1} + (-1)^N c_{j-1}\right] ,& \quad \text{ for } j = 1;\\ \\
      - \sqrt{\Gamma_{\rm eff}} \; c_j \left[ (-1)^N  c_{j+1} -  c_{j-1}\right] ,& \quad \text{ for } j = L \; \text{and $j+1=1$}.
    \end{array}\right.
  \end{align}
\end{subequations}
The correction to $H_1$ corresponds to nearest-neighbor hopping, but the sign of the hopping amplitude depends on the parity of the number of particles considered. As a consequence, the quantized momenta depend on the parity:
\begin{equation}
  c_j = \frac{1}{\sqrt{L}} \sum_k e^{i kj} c_k; \qquad \quad  k = \left\{
  \begin{array}{lll} 
    \frac{2 \pi n}{L}, \quad & n = 1,2, \ldots L, \quad & \text{ for } N \text{ odd};\\ \\
    \frac{2 \pi n}{L}+ \frac{\pi}{L}, \quad & n = 1,2, \ldots L, \quad & \text{ for } N \text{ even};
  \end{array}
  \right. \label{Eq:quantisation}
\end{equation}
In the following we only deal with situations where the initial state has a well-defined parity of the number of particles; since atoms are lost in pairs, the parity remains a well-defined quantum number at all times, so that the term $(-1)^N$ can be taken as a scalar (\textit{i.e.} not an operator). For simplicity we will deliberately neglect some boundary effects, such as the parity dependence of the boundary jump operators, and consider the thermodynamic limit where $L \to + \infty$. The quantisation condition in~\eqref{Eq:quantisation} is crucial for a good description of the exact numerical data (quantum trajectories) with the rate equation that we are developing here.}

\subsection{The rate equation: derivation}
\label{App:Rigorous}

We are interested in the operator $n_k = c_k^\dagger c_k$ and in the equation of motion
for its expectation value $n_k(t) = \langle c_k^\dagger c_k \rangle_t = \text{tr} \left[ c_k^\dagger c_k \rho(t) \right]$. 
When the time evolution is governed by a Lindblad master equation, the following relation holds for any operator $A$ (we use the same notation as in the main text):
\begin{equation}
  \frac{\rm d}{\mathrm d t}  \langle A \rangle_t = + \frac{i}{\hbar}
  \big\langle \left[ H_1+H_2, A \right] \big\rangle_t + 
 \frac{1}{2} \sum_j \left \langle 
  L_j^\dagger [A, L_j] + \mathrm{h.c.}
  \right \rangle_t .
\end{equation}
This relation is straightforwardly derived from the master equation, making repeated use of the cyclic invariance of the trace operation.

For our specific problem and for $A=c_k^\dagger c_k$, this expression can be further simplified. First, we observe that $[H_1, n_k]=0$, such that this Hamiltonian contribution disappears. Second, we have
\begin{subequations}
  \begin{eqnarray}
 L_j^\dagger
 \left[ n_k , L_j \right] & = & - \frac{4 \Gamma_{\rm eff}}{L^2} \sum_{q,q',k'} e^{i (k+q-k'-q')j} \sin(k') \big[ \sin(k)-\sin(q)\big] c^\dagger_{q'} c^\dagger_{k'} c_k c_q.
  \end{eqnarray}
\end{subequations}
Summing over the sites $j$, we obtain
\begin{equation}
  \sum_j L_j^\dagger
  \left[ n_k , L_j \right] = - \frac{4 \Gamma_{\rm eff}}{L} \sum_{q,q',k'} \sin(k') \big[ \sin(k)-\sin(q) \big] \,
  \delta_{k+q,k'+q'}\; c^\dagger_{q'} c^\dagger_{k'} c_k c_q .
\end{equation}
Moving to expectation values, we obtain:
\begin{align}
 \frac{\rm d}{\mathrm d t}  n_k(t) =+ \frac{i}{\hbar}\langle [H_2, n_k ]\rangle
 - \frac{4 \Gamma_{\rm eff}}{L} \sum_{q,q'} \sin(k') \big[ \sin(k)-\sin(q) \big] \langle c^\dagger_{q'} c^\dagger_{k+q-q'} c_k c_q \rangle_t.
 \end{align}
The evolution of one-body operators is thus coupled to two-body operators, the first equation in the familiar Bogoliubov-Born-Green-Kirkwood-Yvon hierarchy typical of many-body problems. 

In order to break the hierarchy and to bring the latter equation to a usable form, we now make our key approximation. We assume that the density matrix is represented by a time-dependent generalized-Gibbs-ensemble (tGGE):
\begin{equation}
  \rho(t) = \frac{1}{\mathcal Z(t)} \exp \bigg[ -\sum_k \mu_k (t) n_k \bigg] , \qquad
  {\mathcal Z(t)} = \prod \left[ 1+ e^{-\mu_k (t)} \right].
  \label{Eq:TGGE:SM}
\end{equation}
For the motivation of this approximation, see also Sec.~\ref{Sec:tGGE:SM}.
This Gaussian quantum state satisfies Wick's theorem, $
  \langle c_z^\dagger c_w^\dagger c_k c_q \rangle_t = 
  \langle c_z^\dagger c_q \rangle_t \langle c_w^\dagger c_k \rangle_t  -
  \langle c_z^\dagger c_k \rangle_t \langle c_w^\dagger c_q \rangle_t$,
and factorization in momentum space, $\langle c^\dagger_k c_q \rangle_t = \delta_{k,q} n_k(t)$. As a result, we have
\begin{equation}
  \langle c_{q'}^\dagger c_{k+q-q'}^\dagger c_k c_q \rangle_t = 
  \left(\delta_{q,q' }-\delta_{k,q'} \right) n_k(t) \, n_q(t), \end{equation}
and $  \langle c_z^\dagger c_w^\dagger c_k c_q \rangle_t - \langle c_q^\dagger c_k^\dagger c_w c_z \rangle_t = 0$.
Using this relation, we are able to break the hierarchy of equations of motions to first order.
Moreover, the Hamiltonian part of the dynamics that depends on $ H_2$ can also be simplified because $[ n_k, \rho(t)]=0$. Indeed, invoking the cyclic property of the trace:
\begin{equation}
 \langle [H_2,  n_k] \rangle_t = \text{tr}\big[  [H_2,  n_k ] \, \rho \big]=\text{tr}\big[ H_2\, [ n_k, \rho ]\big]=0.
\end{equation}

We finally obtain:
\begin{align}
 \frac{\rm d}{\mathrm d t}  n_k(t) =& - \frac{4\Gamma_{\rm eff}}{L} \sum_q \big(\sin (k)-\sin(q) \big)^2  n_k(t)   n_q(t),
 \label{Eq:RateEquation:SM}
 \end{align}
which is the rate equation presented in the main text.
If $\sum_q \sin(q) n_q(t) = 0$, for instance when the momentum distribution is inversion-symmetric (a property that is preserved during the time evolution), the rate equation can be further simplified: 
\begin{equation}
 \frac{\rm d}{\mathrm d t}  n_k(t) = - \frac{4\Gamma_{\rm eff}}{L}  \sum_q \left[ \sin^2(k)+\sin^2 (q) \right] n_k(t)n_q(t).
 \label{Eq:Rate:Eq:SumRule}
\end{equation}

\subsection{On the neglection of $H_2$}
\label{Sec:No:H2}

\begin{figure}[t]
\includegraphics[width=0.66\textwidth]{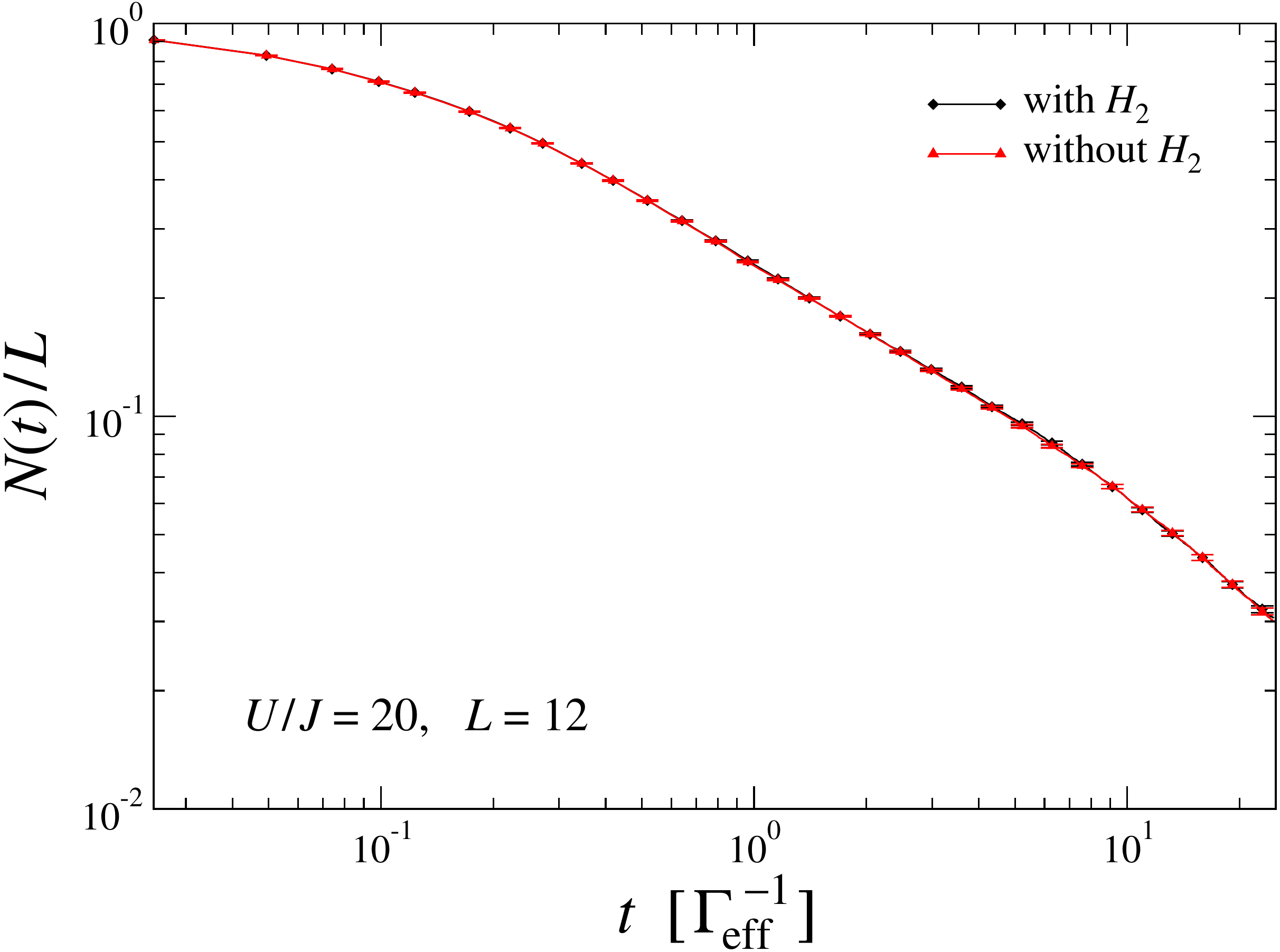}
 \caption{Numerical simulations performed using quantum trajectories of the master equation~\eqref{Eq:SUPP:MEQ}. The red curve has been obtained neglecting $H_2 $ in order to show that it does not contribute significantly to the dynamics.}
 \label{Fig:Supp:NoH2} 
\end{figure} 

The derivation of the rate equations that we have just presented shows that $H_2$ does not play an important role in the dynamics;
in this Section we verify this statement with an independent method.
In Fig.~\ref{Fig:Supp:NoH2} we present some numerical simulations of the master equation~\eqref{Eq:SUPP:MEQ}  performed using quantum trajectories. The black curve is the standard one, whereas the red has been obtained by deliberately neglecting $H_2$ from the simulations.
To all purposes, the comparison of the curves shows negligible differences, that are of the order of the differences with the rate equation.
This plot makes apparent that $H_2$ does not contribute to the dynamics described in this article.

\section{Long-time behaviour of the rate equation}

In this Section we discuss the asymptotic behavior of the solution of the rate equation~(5) in the main text
[cf.~Eq.~\eqref{Eq:RateEquation:SM} in this Supplementary Material]. 
We first note that $\sum_k \sin(k)n_k(t)=0$ at every time.
Indeed, this property holds by assumption at $t=0$ where $n_k(0)=1$. Since the dynamics is invariant under exchange of $k \to -k$, it therefore preserves the property $\sum_k \sin(k)n_k(t)=0$.

We first introduce the total atom number $N(t)=\sum_k n_k$ and the average filling factor $n(t)=N(t)/L$. 
After rewriting the rate equation~\eqref{Eq:Rate:Eq:SumRule} as: 
\begin{equation*}
  \frac{\mathrm d n_k(t)}{\mathrm d t} = - 4 \Gamma_{\rm eff} \left[ 
    \bigg( \frac{1}{L}\sum_q \sin^2(q) n_q(t) \bigg) n_k(t)+ \sin^2(k) n_k(t) n(t)
    \right],
\end{equation*}
 we obtain that
\begin{equation}
  \frac{\mathrm d n(t)}{\mathrm dt} = -8\Gamma_\mathrm{eff} \bigg( \sum_{q} \sin^2(q) {n}_{q}(t) \bigg) n(t).
\end{equation}
We can thus rewrite the rate equation using a scaled time $\tau=4\Gamma_\mathrm{eff} t$ and the notation $\dot{f}=\mathrm df/\mathrm d\tau$,
\begin{equation}
  \dot{{n}}_k(\tau) = \frac{ \dot{{n}}(\tau) {n}_k(\tau)}{2 {n}(\tau)} - \sin^2(k) \,{n}(\tau){n}_k(\tau).
\end{equation}
Changing variable to
\begin{equation}
f_k(\tau) = \frac{{n}_k(\tau)}{\sqrt{{n}(\tau)}},
\end{equation}
one readily finds that the function $f_k$ obeys the differential equation
\begin{equation}
  \frac{\dot{f}_k(\tau)}{f_k(\tau)} = -{n}(\tau) \sin^2(k).
\end{equation}
Since the right-hand-side obeys separation of variables, the solution is of the form
\begin{equation}
  f_k(\tau)  = f_k(0) e^{-\sin^2(k) g(\tau)},
\end{equation}
where the primitive
\begin{equation}
  g(\tau)= \int_0^\tau {n}(t')\, {\rm d}t'
\end{equation}
is still unknown at this stage. 

We now promote $k$ to a continuous variable, so that the relation $N(\tau)=\sum_k n_k(\tau)$ now reads $N(\tau) = \frac{L}{2 \pi} \int_{-\pi}^{+\pi} \!\! n_k(\tau ){\rm d}k$. Assuming that the initial state is the ground state of a free-fermion Hamiltonian, namely a zero-temperature Fermi-Dirac distribution with $f_k(0)=1/\sqrt{n_0}$ for $k\in[- \pi n_0, \pi n_0]$ and zero otherwise, we obtain for the density $n(\tau)=N(\tau)/L$,
\begin{equation}
  \sqrt{n(\tau)} =
  \frac{1}{2\pi} \int_{-\pi }^{\pi} \!\! f_k(\tau) {\rm d} k = 
  \frac{1}{2\pi \sqrt{n_0}} \int_{-\pi n_0}^{\pi n_0} \!\! e^{- \sin^2(k) g(\tau)} {\rm d} k .
  \label{Eq:Fund:Fill}
\end{equation}
In the rest of the Section we discuss the long-time limit of the system for different initial conditions.

\subsection{Band insulator with initial density $n_0=1$}

When starting from a band insulator with $n=1$, one also has $f_k(0)=1$. The normalization of the distribution yields
\begin{equation}
  \sqrt{n(\tau)}=\frac{1 }{2\pi} \int_{-\pi}^\pi e^{- \sin^2(k) g(\tau)} \mathrm dk
  = e^{-\frac{g(\tau)}{2}} \, I_0 \bigg[ \frac{g(\tau)}{2} \bigg],
 \label{eq:crappyrelation}
\end{equation}
where $I_0$ is a modified Bessel function of the first kind. 
Although this equation does not seem analytically solvable, an asymptotic analysis gives interesting insight into the long-time
dynamics of the system. Let us focus on physically relevant solutions where $g$ is a monotonically increasing function (so that the atom number actually decreases in time). For long enough times, we can then use the asymptotic expansion $ I_0(x) \sim e^x/\sqrt{2\pi x}$ for $x \to +\infty$. The relation (\ref{eq:crappyrelation}) then becomes $  \pi {n}(\tau) g(\tau) \approx 1$. Considering that $\dot g = n(\tau)$, we can
write the differential equation as
\begin{equation}
\dot g(\tau) g(\tau) = 1/\pi.
\end{equation}
The solution is
\begin{equation}
  \label{eq:nbarsqrt}
  g(\tau)=\sqrt{c_0+\frac{2}{\pi}\tau},
\end{equation}
where $c_0$ is a constant.

We thus find that the solution is universal at long times and behaves asymptotically as
\begin{equation}
  \label{eq:nkas}
  n(\tau)\sim\frac{1}{\sqrt{2\pi\tau}};
  \qquad 
  n_k(\tau) \sim \frac{1}{(2\pi\tau)^{1/4}} e^{-\sin^2(k) \sqrt{\frac{2\tau}{\pi}}}.
\end{equation}
This asymptotic form is valid for $n \ll 1$, or equivalently $\tau \gg 1/(2\pi)$. Note that $g(\tau)$ diverges monotonically in the long-time limit and we thus verify a posteriori the assumption.
Interestingly, this solution predicts that the asymptotic distribution when $t \to + \infty$
is made of two momenta $k=0,\pi$, whose populations decay as $t^{-\frac{1}{4}}$
instead as $\exp[-\sqrt{t / z_k}]$, where $z_k$ is a $k$-dependent time scale.
We recognize in Eq.~\eqref{eq:nkas} the Eq.~(6) of the main text.

\subsection{Half filling case ($n_0=0.5$)}

The case of half filling is mathematically very similar to the previous one. We find
\begin{equation}
  \sqrt{n(\tau)}=\frac{1}{\sqrt 2\pi} \int_{-\frac{\pi}2}^{\frac{\pi}2} e^{- \sin^2(k) g(\tau)} \mathrm dk
  = \frac 1{\sqrt 2} e^{-\frac{g(\tau)}{2}} I_0 \bigg[ \frac{g(\tau)}{2} \bigg].
\end{equation}
In the long-time limit, we obtain $\dot g(\tau) g(\tau) = 1/(2 \pi)$ and
\begin{equation}
  n(\tau)\sim\frac{1}{\sqrt{4\pi\tau}};
  \qquad 
  n_k(\tau) \sim \frac{1}{(4\pi\tau)^{1/4}} e^{-\sin^2(k) \sqrt{\frac{\tau}{\pi}}}, \;\; k \in \left[ - \frac{\pi}{2}, + \frac{\pi}{2}\right].
\end{equation}
The result is thus qualitatively very similar to the case of an initial band insulator $n=1$.

\subsection{Low filling case ($n_0\ll 1$)}

We start again from relation~\eqref{Eq:Fund:Fill}. In this case, the integration
is restricted to a small neighbourhood around $k = 0$, and we perform a Taylor expansion of $\sin(k)$. We find
\begin{equation}
  \sqrt{n(\tau)} = \frac{1}{2\pi} \int_{-\pi }^{\pi} f_k(\tau) {\rm d} k = 
  \frac{1}{2\pi \sqrt{n_0}} \int_{-\pi n_0}^{\pi n_0} e^{- k^2 g(\tau)} {\rm d} k .
  \label{Eq:Low:Fill:Sub}
\end{equation}
The integrand takes substantial values for $ \vert k \vert \lesssim \frac{1}{ \sqrt{g(\tau)}}$. At long times such that $g(\tau) \gg 1/(\pi n_0)^2$, we can extend the integration limits to infinity,
\begin{equation}
  \sqrt{n(\tau)} \approx \frac{1}{2 \pi \sqrt{n_0}} \int_{-\infty}^{+ \infty}e^{- k^2 g(\tau)} {\rm d} k = \frac{1}{\sqrt{4 \pi n_0 g(\tau)}}.
\end{equation}
We thus obtain $\dot g(\tau) g(\tau) = 1/(4 \pi n_0)$, yielding
\begin{equation}
  n(\tau)\sim\frac{1}{\sqrt{8\pi n_0\tau}};
  \qquad 
  n_k(\tau) \sim \frac{1}{(8\pi n_0\tau)^{1/4}} e^{-\sin^2(k) \sqrt{\frac{\tau}{2\pi n_0}}}, \; \; k \in \left[ - \pi n_0, \pi n_0 \right].
\end{equation}

\subsection{An analytical formula for $N(t)$}

\begin{figure}[t]
 \includegraphics[width=0.66\textwidth]{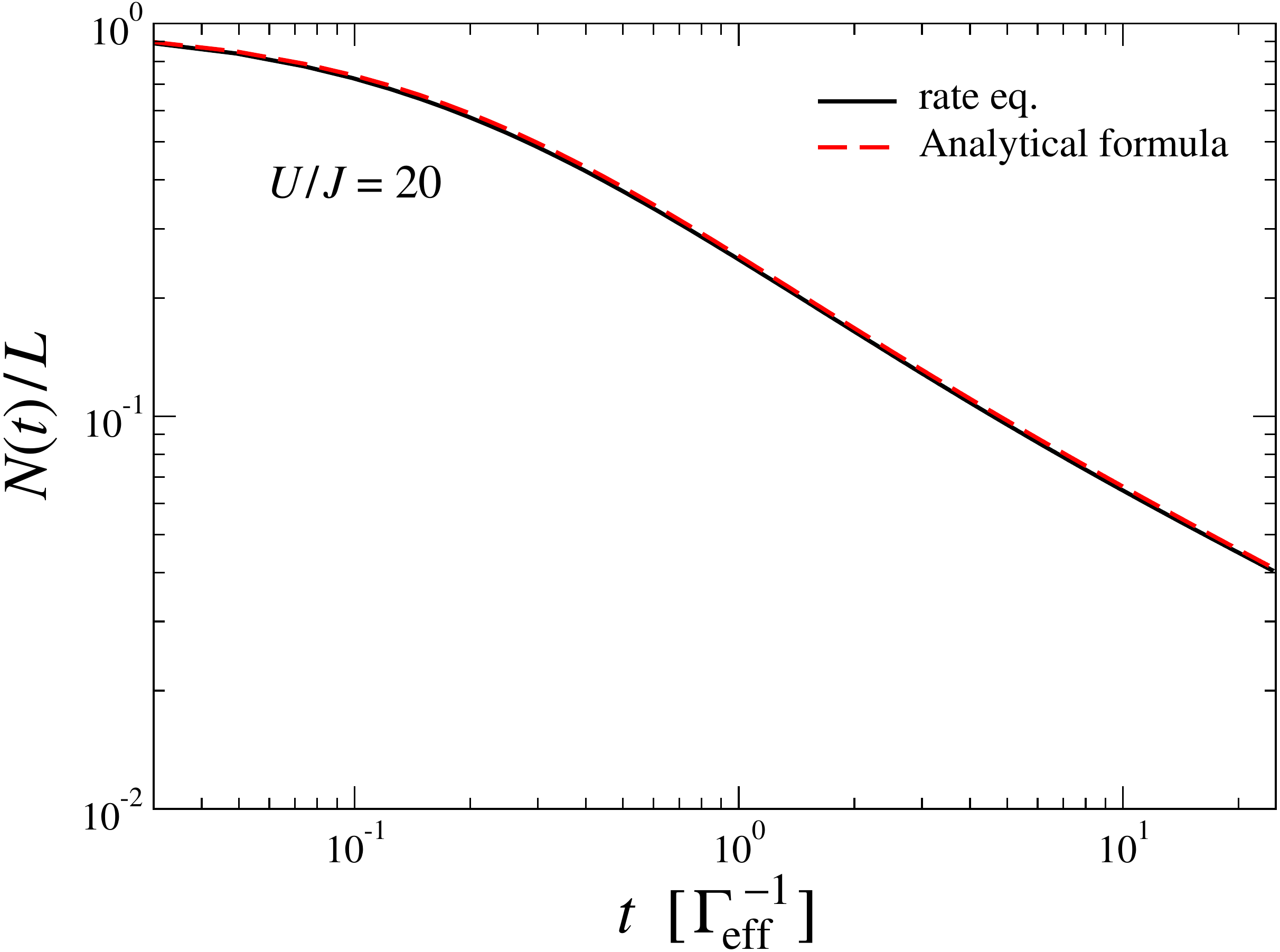}
 \caption{Comparison between the population density $N(t)/L$ obtained with a numerical solution of the rate equations and the analytical formula proposed in Eq.~\eqref{Eq:Supp:Anal}.}
 \label{Fig:Analytical}
\end{figure}

In the main text, the rate equation has been solved using numerical techniques. Although these simulations are not difficult to reproduce, this might make the fit of experimental data quite complicated.
We present here the following analytical formula:
\begin{equation}
 n(\tau) = \frac{\sqrt{1+ \frac{\tau}{2 \pi}}}{1+\tau}; 
 \label{Eq:Supp:Anal}
\end{equation}
for which we do not have an analytical proof but that reproduces the numerical data with a very good accuracy, see Fig.~\ref{Fig:Analytical}.

This analytical formula possesses the correct behaviour at long and short time. At short time, it is approximated by:
\begin{equation}
 n(\tau) \sim 1- \left(1-\frac{1}{4 \pi} \right) \tau, \qquad \tau \ll 1;
\end{equation}
and given that $1/(4 \pi) \sim 0.078...$ there is no appreciable difference with the short-time behaviour proposed by the mean-field solution $N_0(\tau)/ L = (1+ \tau)^{-1} \sim 1-\tau$.
At long times, instead we obtain:
\begin{equation}
 n(\tau) \sim \frac{1}{\sqrt{2 \pi}} \frac{1}{\sqrt \tau}, \qquad \tau \gg 1;
\end{equation}
that coincides with Eq.~\eqref{eq:nkas} derived above.

\section{Momentum distribution functions}

\begin{figure}
\centering
\includegraphics[width=0.66\textwidth]{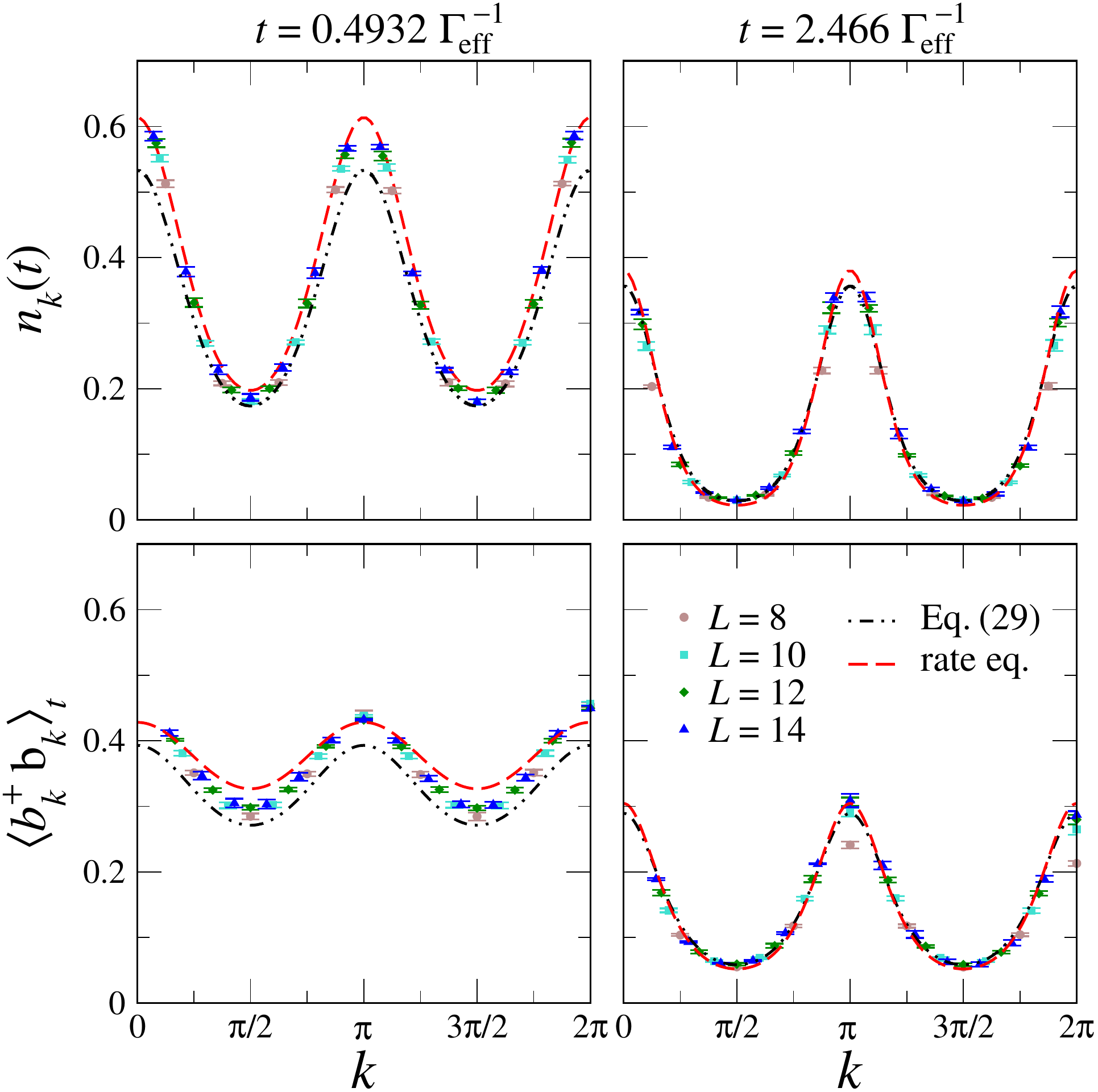} 
\caption{Momentum distribution functions: comparison of data obtained using the quantum trrajectories and different analytical approaches (see text and legend). Different columns are related to different times. In the upper panels we discuss the fermionic momentum distribution function, whereas in the lower panels we discuss the bosonic one.}
\label{Fig:SupMat:1}
\end{figure}

\subsection{Accuracy of the rate equations for the momentum distribution functions}

In this Section we present more data on the accuracy of the calculation of the momentum distribution functions using the rate equations. In Fig.~\ref{Fig:SupMat:1} we present four panels; 
in the upper panels we focus on fermionic momentum distribution function $n_k(t)$ and in the lower panels on bosonic momentum distribution functions $\langle b_k^\dagger b_k \rangle_t$; we consider two different times, $t=0.4932 \Gamma_{\rm eff}^{-1}$ in the left column and $t=2.466 \Gamma_{\rm eff}^{-1}$ in the right column. 
In each panel we compare data obtained (i) using quantum trajectories for different sizes up to $L=14$, (ii) using the formula~\eqref{eq:nkas} in the main text corresponding to a long-time approximation, and (iii) using the rate equation.
The bosonic $\langle b_k^\dagger b_k \rangle_t$ are obtained from the fermionic ones using the methods explained in Ref.~\cite{Gangardt_2006}.
The agreement between the rate equation and the quantum trajectories is very good and increases with time. We observe that formula~\eqref{eq:nkas} from the main text, derived in the long-time limit, gives a satisfactory description of the data only for the panels on the right.

\begin{figure}[t!]
 \includegraphics[height=0.37\textwidth]{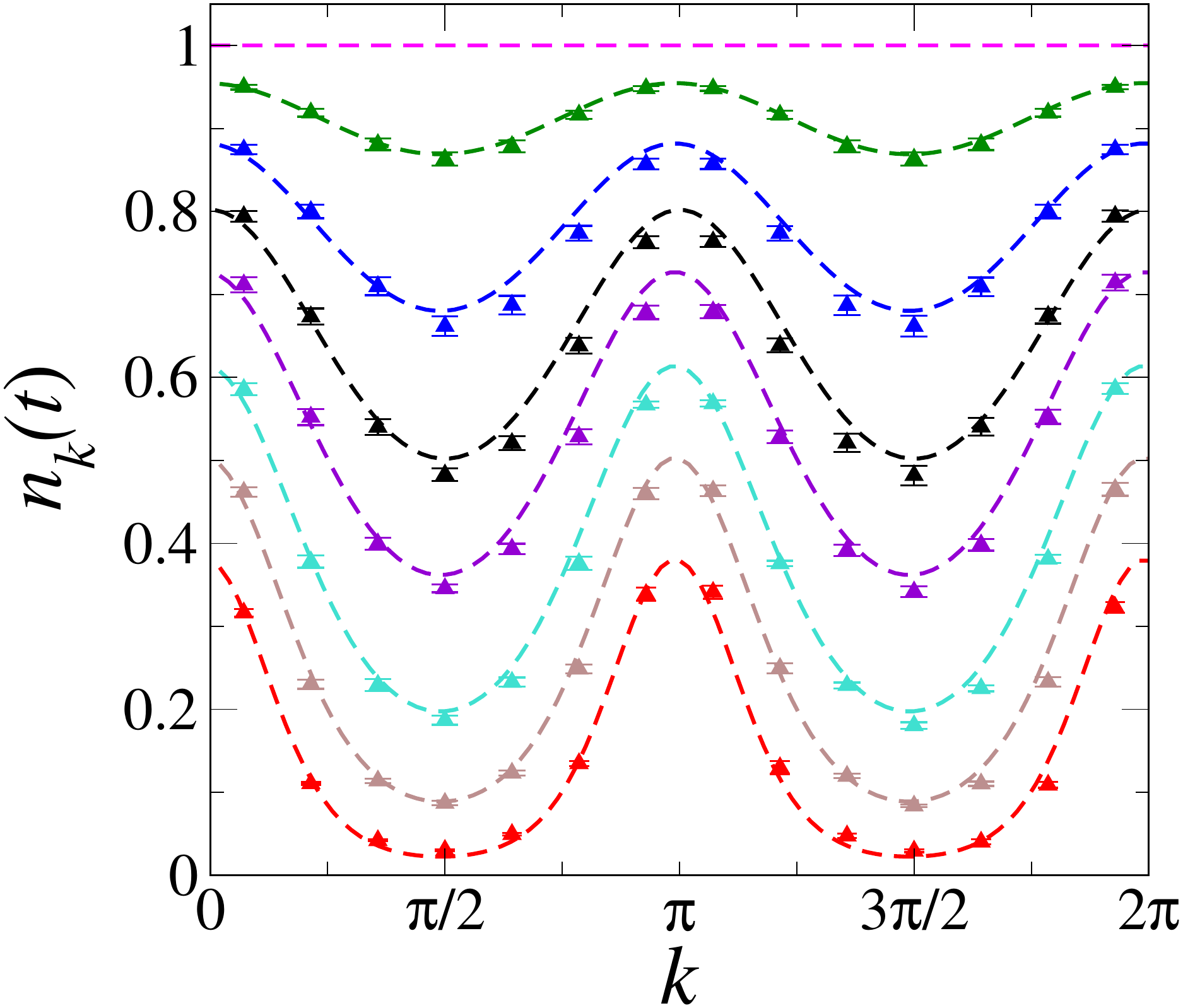}
\hfill
 \includegraphics[height=0.37\textwidth]{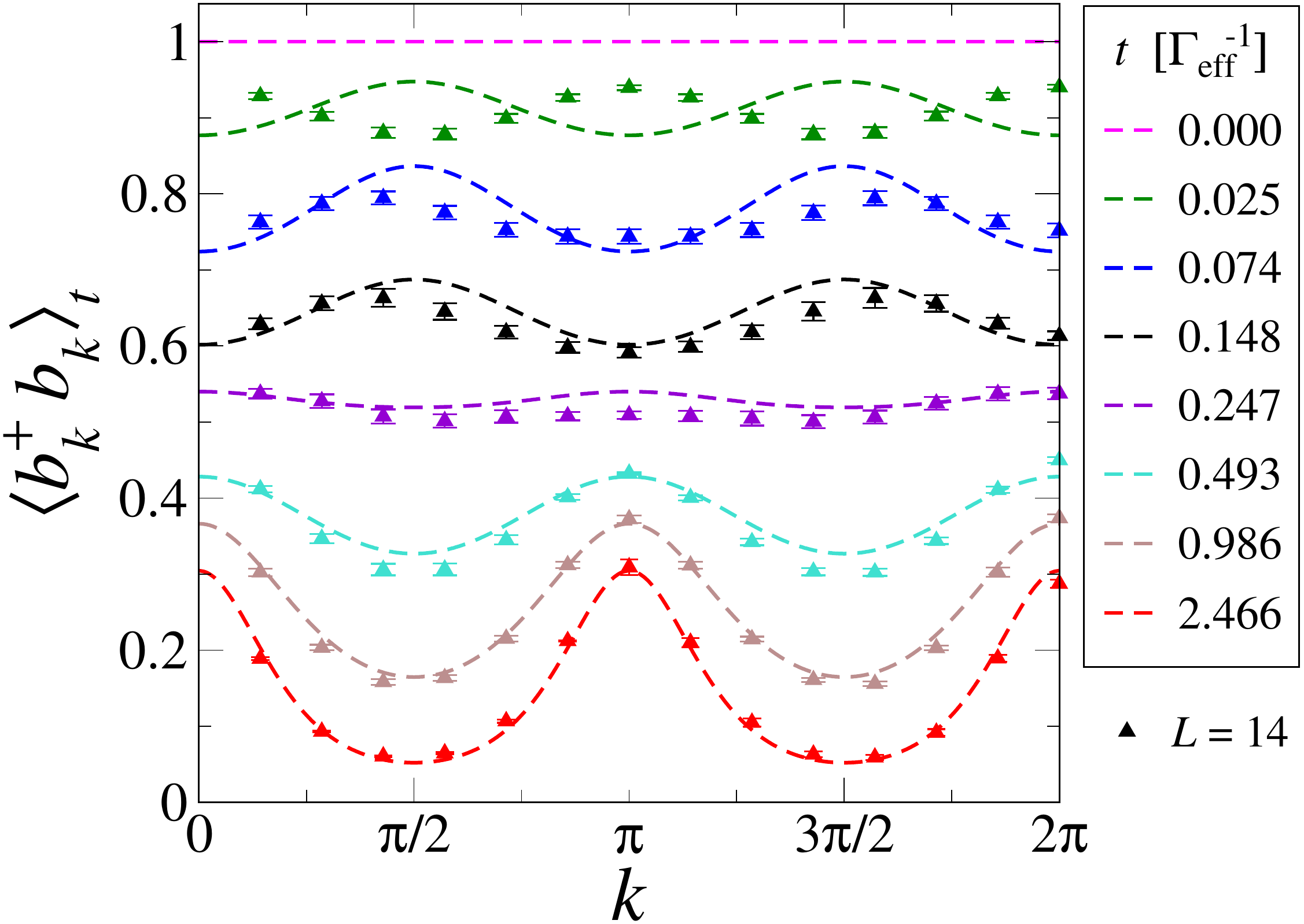}
 \caption{Fermionic (top) and bosonic (bottom) momentum distribution functions at different times. Dashed lines are the rate-equation predictions; triangles represent the results obtained with quantum trajectories for $L=14$. In general, fermionic data are well-described by the rate equations whereas the bosonic ones have a worst agreement at short times.}
 \label{Fig:ShortTimes}
\end{figure}

In the main text we comment on the fact that our rate equation does not properly describe $\langle b_k^\dagger b_k \rangle_t$ at very short times $t < 0.1 \Gamma_{\rm eff}^{-1}$. In Fig.~\ref{Fig:ShortTimes} we show a comparison between the quantum trajectories at $L=14$ and the rate equation for $\langle b_k^\dagger b_k \rangle_t$; our statement can be easily verified.
In order to understand this discrepancy, we have first computed $n_k(t)$ at all corresponding times, and have observed a good agreement between rate-equation data and quantum-trajectory simulations (see Fig.~\ref{Fig:ShortTimes}); this cannot be the reason of the mismatch.

We can thus explain the mismatch by observing that $\langle b_k^\dagger b_k \rangle_t$, which is computed using the method detailed in Ref.~\cite{Gangardt_2006}, also depends on off-diagonal momentum correlators: $\text{Tr}[c_k^\dagger c_{k'} \rho(t)]$.
We have checked numerically that these correlators are small with respect to the actual value of $n_k(t)$, which is of order $1$, but are large compared to the $k$-dependence of $n_k(t)$ at short times. Indeed, the position of the peaks in $\langle b_k^\dagger b_k \rangle_t$ does not depend on the average value $\sum_k n_k(t) / L$, which only determines an average offset, but only on $n_k(t)-\sum_q n_q(t) / L$. If this is correct, neglecting the comparable values of $\text{Tr}[c_k^\dagger c_{k'} \rho(t)]$ introduces an important error. As time increases, these terms become negligible (after meeting a maximum at $t \sim 1/(4 \Gamma_{\rm eff})$) and the $k$ dependence of $n_k(t)$ more pronounced; this explains the improved agreement on $\langle b_k^\dagger b_k \rangle_t$ between the rate equations and the quantum trajectories.


\section{Time-dependent generalized Gibbs ensemble}
\label{Sec:tGGE:SM}

\subsection{The general theory}

The derivation of the rate equation proposed in Sec.~\ref{App:Rigorous} can be framed within the more general derivation of the tGGE proposed in Ref.~\cite{Lange_2018} and briefly recalled at the end of the main text. The idea goes as follows. We consider a dissipative dynamics and a Lindbladian composed of a strong and of a weak part:
\begin{equation}
 \frac{\rm d}{\mathrm d t} \rho(t) = \mathcal L_0[\rho(t)]+ \epsilon \mathcal L_1[\rho (t)], \quad \epsilon \ll 1.
 \label{Eq:Perturbative}
\end{equation}
We furthermore assume that the most relevant term only describes a Hamiltonian dynamics, ruled by $H_0$.
We wish to find a solution to this problem in the limit $\epsilon \to 0^+$. We divide Eq.~\eqref{Eq:Perturbative} by $\epsilon$ and introduce the rescaled time $\tau = \epsilon t$:
\begin{equation}
 \frac{\rm d}{\mathrm d \tau} \rho_\epsilon(\tau) = \frac{1}{\epsilon} \mathcal L_0[\rho_\epsilon(\tau)]+ \mathcal L_1[\rho_\epsilon (\tau)]
 \label{Eq:Perturbative:2}
\end{equation}
and perform the expansion
\begin{equation}
 \rho_\epsilon (\tau ) = \rho^{(0)}(\tau)
 + \epsilon \rho^{(1)}(\tau) + \epsilon^2 \rho^{(2)}(\tau) + O(\epsilon^3).
 \label{Eq:rho:expansion}
\end{equation}
Our goal is to determine $\rho^{(0)}(\tau)$, that is the $\lim_{\epsilon \to 0^+} \rho_\epsilon(\tau)$; 
this will give us access to the leading properties of the solution of equation~\eqref{Eq:Perturbative} when $\epsilon$ is small via the substitution: $\rho(t) = \rho^{(0)}(\epsilon t)$.

We substitute Eq.~\eqref{Eq:rho:expansion} into Eq.~\eqref{Eq:Perturbative:2} and compare order by order. 
The leading term of order $\epsilon^{-1}$ returns the equation:
\begin{equation}
 \mathcal L_0[\rho^{(0)}(\tau)] = 0 .
\end{equation}
As such, the quantum state $\rho^{(0)}(\tau)$ lives in the space of the stationary states of the Hamiltonian, and is diagonal in the basis of its eigenstates (apart from possible degeneracies). The tGGE approximation says that, in order to describe the long-time dynamics, it is enough to assume a simple GGE structure:
\begin{equation}
\label{Eq:TGGE:SM2}
  \rho^{(0)}(\tau) = \frac{1}{\mathcal Z(\tau)} e^{- \sum_k \mu_k(\tau) I_k},
\end{equation}
where the $I_q$ are the conserved quantities of $H_0$ and the associated  Lagrange multiplier $\mu_q(\tau)$ depends on time.
This choice is clearly more restrictive than considering a generic (time-dependent) stationary state.
 
The next order $\epsilon^0$ determines the dynamics of the $\mu_k(\tau)$. We do not re-derive it here but we simply take the result presented in Ref.~\cite{Lange_2018}:
\begin{equation}
 \frac{\mathrm d \mu_k(\tau)}{\mathrm d \tau} = 
 - \sum_q \left[ \chi^{-1}(\tau)\right]_{kq} \text{Tr}\big[ \mathcal L_1^*[I_q] \rho^{(0)}(\tau) \big],
 \qquad 
 \chi_{kq}(\tau) = \langle I_k I_q \rangle_\tau - \langle I_k \rangle_\tau \langle I_q \rangle_\tau.
\end{equation}

\subsection{Our case}

We frame the master equation~\eqref{Eq:SUPP:MEQ} in the form of Eq.~\eqref{Eq:Perturbative}:
\begin{equation}
 \mathcal L_0[\rho(t)] = - \frac{i}{\hbar} [H_1, \rho(t)];
 \qquad \mathcal L_1[\rho(t)] = - \frac{i}{\hbar} [
 H_2,\rho(t) \big] + 
  \sum_j \left[ L_j \rho(t) L_j^\dagger - \frac{1}{2} \left\{ L_j^\dagger L_j, \rho(t) \right\} \right].
\end{equation}
Note that $\mathcal L_1$ is multiplied by the small parameter $J^2 / \hbar^2 \gamma^2$, although this is not put explicitly in evidence.
The conserved quantities of $H_1$ are $I_k \equiv n_k$, so that $\rho^{(0)}(\tau)$ has exactly the form of Eq.~\eqref{Eq:TGGE:SM2} used in the previous derivation. 

In order to derive the dynamics of the different $\mu_k(\tau)$, we observe that:
\begin{subequations}
\begin{align}
  \text{Tr}[\mathcal L_1^*[I_k] \rho^{(0)}(\tau)] &= \Big\langle \sum_j L_j^\dagger \left[n_k, L_j \right]\Big\rangle
  = - \frac{4 \Gamma_{\rm eff}}{L} \sum_q \big[ \sin(k)-\sin(q) \big]^2 n_k (\tau) n_q(\tau); \\
 \chi_{kq}(\tau) &=  n_k(\tau) \big[ n_k(\tau)-1 \big] \delta_{kq} .
\end{align}
\end{subequations}
We obtain:
\begin{equation}
 \frac{\mathrm d \mu_k(\tau)}{\mathrm d \tau} = -
 \frac{1}{n_k(\tau)(n_k(\tau)-1)} \left( \frac{4 \Gamma_{\rm eff}}{L } \right) 
 \sum_q \big[ \sin(k)-\sin(q) \big]^2 n_k (\tau) n_q(\tau).
 \label{Eq:46}
\end{equation}
In order to verify the correctness of this equation, it is useful to make explicit the relation between $\mu_k(\tau)$, the Lagrange multiplier, and the expectation value $n_k(\tau)$:
\begin{equation}
 n_k(\tau) = \frac{\text{Tr}[n_k e^{- \mu_k(\tau) n_k}]}{\text{Tr}[e^{- \mu_k(\tau) n_k}]} = \frac{1}{e^{\mu_k}+1},
 \qquad \Rightarrow \qquad \mu = \log \left( \frac{1}{n_k}-1\right), \quad n_k \in[0,1], \; \mu \in \mathbb R.
\end{equation}
We first compute the derivative of $\mu$ with respect to $\tau$ and obtain:
\begin{equation}
 \frac{\mathrm d \mu_k(\tau)}{\mathrm d \tau} =  \frac{1}{n_k(\tau) \big[n_k(\tau)-1\big]} \frac{\mathrm d n_k(\tau)}{\mathrm d \tau}.
\end{equation}
Substituting this expression in Eq.~\eqref{Eq:46} we obtain the rate equation in Eq.~(5) in the main text,
which confirms the validity of this approach, whose essential merit is that of providing
a systematic framework for deriving these dynamical equations.

We conclude this paragraph by providing a closed expression for the rate equation
in terms of the $\mu_k(\tau)$ coefficients:
\begin{equation}
  \frac{\mathrm d \mu_k(\tau)}{\mathrm d \tau} =
  \frac{4 \Gamma_{\rm eff}}{L} \sum_q \big[ \sin(k)-\sin(q) \big]^2 \:
  \frac{e^{-\mu_k(\tau)}+1}{e^{\mu_q(\tau)}+1}.
\end{equation}
At a first sight, this equation does not seem easier to solve analytically than the rate equations written previously.


\begin{thebibliography}{58}%
\makeatletter
\providecommand \@ifxundefined [1]{%
 \@ifx{#1\undefined}
}%
\providecommand \@ifnum [1]{%
 \ifnum #1\expandafter \@firstoftwo
 \else \expandafter \@secondoftwo
 \fi
}%
\providecommand \@ifx [1]{%
 \ifx #1\expandafter \@firstoftwo
 \else \expandafter \@secondoftwo
 \fi
}%
\providecommand \natexlab [1]{#1}%
\providecommand \enquote  [1]{``#1''}%
\providecommand \bibnamefont  [1]{#1}%
\providecommand \bibfnamefont [1]{#1}%
\providecommand \citenamefont [1]{#1}%
\providecommand \href@noop [0]{\@secondoftwo}%
\providecommand \href [0]{\begingroup \@sanitize@url \@href}%
\providecommand \@href[1]{\@@startlink{#1}\@@href}%
\providecommand \@@href[1]{\endgroup#1\@@endlink}%
\providecommand \@sanitize@url [0]{\catcode `\\12\catcode `\$12\catcode
  `\&12\catcode `\#12\catcode `\^12\catcode `\_12\catcode `\%12\relax}%
\providecommand \@@startlink[1]{}%
\providecommand \@@endlink[0]{}%
\providecommand \url  [0]{\begingroup\@sanitize@url \@url }%
\providecommand \@url [1]{\endgroup\@href {#1}{\urlprefix }}%
\providecommand \urlprefix  [0]{URL }%
\providecommand \Eprint [0]{\href }%
\providecommand \doibase [0]{http://dx.doi.org/}%
\providecommand \selectlanguage [0]{\@gobble}%
\providecommand \bibinfo  [0]{\@secondoftwo}%
\providecommand \bibfield  [0]{\@secondoftwo}%
\providecommand \translation [1]{[#1]}%
\providecommand \BibitemOpen [0]{}%
\providecommand \bibitemStop [0]{}%
\providecommand \bibitemNoStop [0]{.\EOS\space}%
\providecommand \EOS [0]{\spacefactor3000\relax}%
\providecommand \BibitemShut  [1]{\csname bibitem#1\endcsname}%
\let\auto@bib@innerbib\@empty
\bibitem [{\citenamefont {Zurek}(2003)}]{Zurek_2003}%
  \BibitemOpen
  \bibfield  {author} {\bibinfo {author} {\bibfnamefont {W.~H.}\ \bibnamefont
  {Zurek}},\ }\href {\doibase 10.1103/RevModPhys.75.715} {\bibfield  {journal}
  {\bibinfo  {journal} {Rev. Mod. Phys.}\ }\textbf {\bibinfo {volume} {75}},\
  \bibinfo {pages} {715} (\bibinfo {year} {2003})}\BibitemShut {NoStop}%
\bibitem [{\citenamefont {Misra}\ and\ \citenamefont
  {Sudarshan}(1977)}]{Misra_1977}%
  \BibitemOpen
  \bibfield  {author} {\bibinfo {author} {\bibfnamefont {B.}~\bibnamefont
  {Misra}}\ and\ \bibinfo {author} {\bibfnamefont {E.~C.~G.}\ \bibnamefont
  {Sudarshan}},\ }\href {\doibase 10.1063/1.523304} {\bibfield  {journal}
  {\bibinfo  {journal} {J. Math. Phys.}\ }\textbf {\bibinfo {volume} {18}},\
  \bibinfo {pages} {756} (\bibinfo {year} {1977})}\BibitemShut {NoStop}%
\bibitem [{\citenamefont {Itano}\ \emph {et~al.}(1990)\citenamefont {Itano},
  \citenamefont {Heinzen}, \citenamefont {Bollinger},\ and\ \citenamefont
  {Wineland}}]{Itano_1990}%
  \BibitemOpen
  \bibfield  {author} {\bibinfo {author} {\bibfnamefont {W.~M.}\ \bibnamefont
  {Itano}}, \bibinfo {author} {\bibfnamefont {D.~J.}\ \bibnamefont {Heinzen}},
  \bibinfo {author} {\bibfnamefont {J.~J.}\ \bibnamefont {Bollinger}}, \ and\
  \bibinfo {author} {\bibfnamefont {D.~J.}\ \bibnamefont {Wineland}},\ }\href
  {\doibase 10.1103/PhysRevA.41.2295} {\bibfield  {journal} {\bibinfo
  {journal} {Phys. Rev. A}\ }\textbf {\bibinfo {volume} {41}},\ \bibinfo
  {pages} {2295} (\bibinfo {year} {1990})}\BibitemShut {NoStop}%
\bibitem [{\citenamefont {Facchi}\ and\ \citenamefont
  {Pascazio}(2002)}]{Facchi_2002}%
  \BibitemOpen
  \bibfield  {author} {\bibinfo {author} {\bibfnamefont {P.}~\bibnamefont
  {Facchi}}\ and\ \bibinfo {author} {\bibfnamefont {S.}~\bibnamefont
  {Pascazio}},\ }\href {\doibase 10.1103/PhysRevLett.89.080401} {\bibfield
  {journal} {\bibinfo  {journal} {Phys. Rev. Lett.}\ }\textbf {\bibinfo
  {volume} {89}},\ \bibinfo {pages} {080401} (\bibinfo {year}
  {2002})}\BibitemShut {NoStop}%
\bibitem [{\citenamefont {Beige}\ \emph
  {et~al.}(2000{\natexlab{a}})\citenamefont {Beige}, \citenamefont {Braun},\
  and\ \citenamefont {Knight}}]{Beige_2000}%
  \BibitemOpen
  \bibfield  {author} {\bibinfo {author} {\bibfnamefont {A.}~\bibnamefont
  {Beige}}, \bibinfo {author} {\bibfnamefont {D.}~\bibnamefont {Braun}}, \ and\
  \bibinfo {author} {\bibfnamefont {P.~L.}\ \bibnamefont {Knight}},\ }\href
  {\doibase 10.1088/1367-2630/2/1/322} {\bibfield  {journal} {\bibinfo
  {journal} {New J. Phys.}\ }\textbf {\bibinfo {volume} {2}},\ \bibinfo {pages}
  {22} (\bibinfo {year} {2000}{\natexlab{a}})}\BibitemShut {NoStop}%
\bibitem [{\citenamefont {Beige}\ \emph
  {et~al.}(2000{\natexlab{b}})\citenamefont {Beige}, \citenamefont {Braun},
  \citenamefont {Tregenna},\ and\ \citenamefont {Knight}}]{Almut_2000}%
  \BibitemOpen
  \bibfield  {author} {\bibinfo {author} {\bibfnamefont {A.}~\bibnamefont
  {Beige}}, \bibinfo {author} {\bibfnamefont {D.}~\bibnamefont {Braun}},
  \bibinfo {author} {\bibfnamefont {B.}~\bibnamefont {Tregenna}}, \ and\
  \bibinfo {author} {\bibfnamefont {P.~L.}\ \bibnamefont {Knight}},\ }\href
  {\doibase 10.1103/PhysRevLett.85.1762} {\bibfield  {journal} {\bibinfo
  {journal} {Phys. Rev. Lett.}\ }\textbf {\bibinfo {volume} {85}},\ \bibinfo
  {pages} {1762} (\bibinfo {year} {2000}{\natexlab{b}})}\BibitemShut {NoStop}%
\bibitem [{\citenamefont {Kempe}\ \emph {et~al.}(2001)\citenamefont {Kempe},
  \citenamefont {Bacon}, \citenamefont {Lidar},\ and\ \citenamefont
  {Whaley}}]{Kempe_2001}%
  \BibitemOpen
  \bibfield  {author} {\bibinfo {author} {\bibfnamefont {J.}~\bibnamefont
  {Kempe}}, \bibinfo {author} {\bibfnamefont {D.}~\bibnamefont {Bacon}},
  \bibinfo {author} {\bibfnamefont {D.~A.}\ \bibnamefont {Lidar}}, \ and\
  \bibinfo {author} {\bibfnamefont {K.~B.}\ \bibnamefont {Whaley}},\ }\href
  {\doibase 10.1103/PhysRevA.63.042307} {\bibfield  {journal} {\bibinfo
  {journal} {Phys. Rev. A}\ }\textbf {\bibinfo {volume} {63}},\ \bibinfo
  {pages} {042307} (\bibinfo {year} {2001})}\BibitemShut {NoStop}%
\bibitem [{\citenamefont {Barreiro}\ \emph {et~al.}(2011)\citenamefont
  {Barreiro}, \citenamefont {Müller}, \citenamefont {Schindler}, \citenamefont
  {Nigg}, \citenamefont {Monz}, \citenamefont {Chwalla}, \citenamefont
  {Hennrich}, \citenamefont {Roos}, \citenamefont {Zoller},\ and\ \citenamefont
  {Blatt}}]{Barreiro_2011}%
  \BibitemOpen
  \bibfield  {author} {\bibinfo {author} {\bibfnamefont {J.~T.}\ \bibnamefont
  {Barreiro}}, \bibinfo {author} {\bibfnamefont {M.}~\bibnamefont {Müller}},
  \bibinfo {author} {\bibfnamefont {P.}~\bibnamefont {Schindler}}, \bibinfo
  {author} {\bibfnamefont {D.}~\bibnamefont {Nigg}}, \bibinfo {author}
  {\bibfnamefont {T.}~\bibnamefont {Monz}}, \bibinfo {author} {\bibfnamefont
  {M.}~\bibnamefont {Chwalla}}, \bibinfo {author} {\bibfnamefont
  {M.}~\bibnamefont {Hennrich}}, \bibinfo {author} {\bibfnamefont {C.~F.}\
  \bibnamefont {Roos}}, \bibinfo {author} {\bibfnamefont {P.}~\bibnamefont
  {Zoller}}, \ and\ \bibinfo {author} {\bibfnamefont {R.}~\bibnamefont
  {Blatt}},\ }\href {\doibase 10.1038/nature09801} {\bibfield  {journal}
  {\bibinfo  {journal} {Nature}\ }\textbf {\bibinfo {volume} {470}},\ \bibinfo
  {pages} {486–491} (\bibinfo {year} {2011})}\BibitemShut {NoStop}%
\bibitem [{\citenamefont {Boulier}\ \emph {et~al.}(2020)\citenamefont
  {Boulier}, \citenamefont {Jacquet}, \citenamefont {Maître}, \citenamefont
  {Lerario}, \citenamefont {Claude}, \citenamefont {Pigeon}, \citenamefont
  {Glorieux}, \citenamefont {Amo}, \citenamefont {Bloch}, \citenamefont
  {Bramati},\ and\ \citenamefont {Giacobino}}]{Boulier_2020}%
  \BibitemOpen
  \bibfield  {author} {\bibinfo {author} {\bibfnamefont {T.}~\bibnamefont
  {Boulier}}, \bibinfo {author} {\bibfnamefont {M.~J.}\ \bibnamefont
  {Jacquet}}, \bibinfo {author} {\bibfnamefont {A.}~\bibnamefont {Maître}},
  \bibinfo {author} {\bibfnamefont {G.}~\bibnamefont {Lerario}}, \bibinfo
  {author} {\bibfnamefont {F.}~\bibnamefont {Claude}}, \bibinfo {author}
  {\bibfnamefont {S.}~\bibnamefont {Pigeon}}, \bibinfo {author} {\bibfnamefont
  {Q.}~\bibnamefont {Glorieux}}, \bibinfo {author} {\bibfnamefont
  {A.}~\bibnamefont {Amo}}, \bibinfo {author} {\bibfnamefont {J.}~\bibnamefont
  {Bloch}}, \bibinfo {author} {\bibfnamefont {A.}~\bibnamefont {Bramati}}, \
  and\ \bibinfo {author} {\bibfnamefont {E.}~\bibnamefont {Giacobino}},\ }\href
  {\doibase 10.1002/qute.202000052} {\bibfield  {journal} {\bibinfo  {journal}
  {Adv. Quantum Technol.}\ ,\ \bibinfo {pages} {2000052}} (\bibinfo {year}
  {2020})}\BibitemShut {NoStop}%
\bibitem [{\citenamefont {Carusotto}\ \emph {et~al.}(2020)\citenamefont
  {Carusotto}, \citenamefont {Houck}, \citenamefont {Koll{\'a}r}, \citenamefont
  {Roushan}, \citenamefont {Schuster},\ and\ \citenamefont
  {Simon}}]{Carusotto_2020}%
  \BibitemOpen
  \bibfield  {author} {\bibinfo {author} {\bibfnamefont {I.}~\bibnamefont
  {Carusotto}}, \bibinfo {author} {\bibfnamefont {A.~A.}\ \bibnamefont
  {Houck}}, \bibinfo {author} {\bibfnamefont {A.~J.}\ \bibnamefont
  {Koll{\'a}r}}, \bibinfo {author} {\bibfnamefont {P.}~\bibnamefont {Roushan}},
  \bibinfo {author} {\bibfnamefont {D.~I.}\ \bibnamefont {Schuster}}, \ and\
  \bibinfo {author} {\bibfnamefont {J.}~\bibnamefont {Simon}},\ }\href
  {\doibase 10.1038/s41567-020-0815-y} {\bibfield  {journal} {\bibinfo
  {journal} {Nat. Phys.}\ }\textbf {\bibinfo {volume} {16}},\ \bibinfo {pages}
  {268} (\bibinfo {year} {2020})}\BibitemShut {NoStop}%
\bibitem [{\citenamefont {S\"oding}\ \emph {et~al.}(1999)\citenamefont
  {S\"oding}, \citenamefont {Gu\'ery-Odelin}, \citenamefont {Desbiolles},
  \citenamefont {Chevy}, \citenamefont {Inamori},\ and\ \citenamefont
  {Dalibard}}]{Soeding_1999}%
  \BibitemOpen
  \bibfield  {author} {\bibinfo {author} {\bibfnamefont {J.}~\bibnamefont
  {S\"oding}}, \bibinfo {author} {\bibfnamefont {D.}~\bibnamefont
  {Gu\'ery-Odelin}}, \bibinfo {author} {\bibfnamefont {P.}~\bibnamefont
  {Desbiolles}}, \bibinfo {author} {\bibfnamefont {F.}~\bibnamefont {Chevy}},
  \bibinfo {author} {\bibfnamefont {H.}~\bibnamefont {Inamori}}, \ and\
  \bibinfo {author} {\bibfnamefont {J.}~\bibnamefont {Dalibard}},\ }\href
  {\doibase https://doi.org/10.1007/s003400050805} {\bibfield  {journal}
  {\bibinfo  {journal} {Appl. Phys. B}\ }\textbf {\bibinfo {volume} {69}},\
  \bibinfo {pages} {257} (\bibinfo {year} {1999})}\BibitemShut {NoStop}%
\bibitem [{\citenamefont {Laburthe~Tolra}\ \emph {et~al.}(2004)\citenamefont
  {Laburthe~Tolra}, \citenamefont {O'Hara}, \citenamefont {Huckans},
  \citenamefont {Phillips}, \citenamefont {Rolston},\ and\ \citenamefont
  {Porto}}]{Tolra_2004}%
  \BibitemOpen
  \bibfield  {author} {\bibinfo {author} {\bibfnamefont {B.}~\bibnamefont
  {Laburthe~Tolra}}, \bibinfo {author} {\bibfnamefont {K.~M.}\ \bibnamefont
  {O'Hara}}, \bibinfo {author} {\bibfnamefont {J.~H.}\ \bibnamefont {Huckans}},
  \bibinfo {author} {\bibfnamefont {W.~D.}\ \bibnamefont {Phillips}}, \bibinfo
  {author} {\bibfnamefont {S.~L.}\ \bibnamefont {Rolston}}, \ and\ \bibinfo
  {author} {\bibfnamefont {J.~V.}\ \bibnamefont {Porto}},\ }\href {\doibase
  10.1103/PhysRevLett.92.190401} {\bibfield  {journal} {\bibinfo  {journal}
  {Phys. Rev. Lett.}\ }\textbf {\bibinfo {volume} {92}},\ \bibinfo {pages}
  {190401} (\bibinfo {year} {2004})}\BibitemShut {NoStop}%
\bibitem [{\citenamefont {Haller}\ \emph {et~al.}(2011)\citenamefont {Haller},
  \citenamefont {Rabie}, \citenamefont {Mark}, \citenamefont {Danzl},
  \citenamefont {Hart}, \citenamefont {Lauber}, \citenamefont {Pupillo},\ and\
  \citenamefont {N\"agerl}}]{Haller_2011}%
  \BibitemOpen
  \bibfield  {author} {\bibinfo {author} {\bibfnamefont {E.}~\bibnamefont
  {Haller}}, \bibinfo {author} {\bibfnamefont {M.}~\bibnamefont {Rabie}},
  \bibinfo {author} {\bibfnamefont {M.~J.}\ \bibnamefont {Mark}}, \bibinfo
  {author} {\bibfnamefont {J.~G.}\ \bibnamefont {Danzl}}, \bibinfo {author}
  {\bibfnamefont {R.}~\bibnamefont {Hart}}, \bibinfo {author} {\bibfnamefont
  {K.}~\bibnamefont {Lauber}}, \bibinfo {author} {\bibfnamefont
  {G.}~\bibnamefont {Pupillo}}, \ and\ \bibinfo {author} {\bibfnamefont
  {H.-C.}\ \bibnamefont {N\"agerl}},\ }\href {\doibase
  10.1103/PhysRevLett.107.230404} {\bibfield  {journal} {\bibinfo  {journal}
  {Phys. Rev. Lett.}\ }\textbf {\bibinfo {volume} {107}},\ \bibinfo {pages}
  {230404} (\bibinfo {year} {2011})}\BibitemShut {NoStop}%
\bibitem [{\citenamefont {Schmidutz}\ \emph {et~al.}(2014)\citenamefont
  {Schmidutz}, \citenamefont {Gotlibovych}, \citenamefont {Gaunt},
  \citenamefont {Smith}, \citenamefont {Navon},\ and\ \citenamefont
  {Hadzibabic}}]{Schmidutz_2014}%
  \BibitemOpen
  \bibfield  {author} {\bibinfo {author} {\bibfnamefont {T.~F.}\ \bibnamefont
  {Schmidutz}}, \bibinfo {author} {\bibfnamefont {I.}~\bibnamefont
  {Gotlibovych}}, \bibinfo {author} {\bibfnamefont {A.~L.}\ \bibnamefont
  {Gaunt}}, \bibinfo {author} {\bibfnamefont {R.~P.}\ \bibnamefont {Smith}},
  \bibinfo {author} {\bibfnamefont {N.}~\bibnamefont {Navon}}, \ and\ \bibinfo
  {author} {\bibfnamefont {Z.}~\bibnamefont {Hadzibabic}},\ }\href {\doibase
  10.1103/PhysRevLett.112.040403} {\bibfield  {journal} {\bibinfo  {journal}
  {Phys. Rev. Lett.}\ }\textbf {\bibinfo {volume} {112}},\ \bibinfo {pages}
  {040403} (\bibinfo {year} {2014})}\BibitemShut {NoStop}%
\bibitem [{\citenamefont {Labouvie}\ \emph {et~al.}(2016)\citenamefont
  {Labouvie}, \citenamefont {Santra}, \citenamefont {Heun},\ and\ \citenamefont
  {Ott}}]{Ott_2016}%
  \BibitemOpen
  \bibfield  {author} {\bibinfo {author} {\bibfnamefont {R.}~\bibnamefont
  {Labouvie}}, \bibinfo {author} {\bibfnamefont {B.}~\bibnamefont {Santra}},
  \bibinfo {author} {\bibfnamefont {S.}~\bibnamefont {Heun}}, \ and\ \bibinfo
  {author} {\bibfnamefont {H.}~\bibnamefont {Ott}},\ }\href {\doibase
  10.1103/PhysRevLett.116.235302} {\bibfield  {journal} {\bibinfo  {journal}
  {Phys. Rev. Lett.}\ }\textbf {\bibinfo {volume} {116}},\ \bibinfo {pages}
  {235302} (\bibinfo {year} {2016})}\BibitemShut {NoStop}%
\bibitem [{\citenamefont {Rauer}\ \emph {et~al.}(2016)\citenamefont {Rauer},
  \citenamefont {Gri\ifmmode~\check{s}\else \v{s}\fi{}ins}, \citenamefont
  {Mazets}, \citenamefont {Schweigler}, \citenamefont {Rohringer},
  \citenamefont {Geiger}, \citenamefont {Langen},\ and\ \citenamefont
  {Schmiedmayer}}]{Rauer_2016}%
  \BibitemOpen
  \bibfield  {author} {\bibinfo {author} {\bibfnamefont {B.}~\bibnamefont
  {Rauer}}, \bibinfo {author} {\bibfnamefont {P.}~\bibnamefont
  {Gri\ifmmode~\check{s}\else \v{s}\fi{}ins}}, \bibinfo {author} {\bibfnamefont
  {I.~E.}\ \bibnamefont {Mazets}}, \bibinfo {author} {\bibfnamefont
  {T.}~\bibnamefont {Schweigler}}, \bibinfo {author} {\bibfnamefont
  {W.}~\bibnamefont {Rohringer}}, \bibinfo {author} {\bibfnamefont
  {R.}~\bibnamefont {Geiger}}, \bibinfo {author} {\bibfnamefont
  {T.}~\bibnamefont {Langen}}, \ and\ \bibinfo {author} {\bibfnamefont
  {J.}~\bibnamefont {Schmiedmayer}},\ }\href {\doibase
  10.1103/PhysRevLett.116.030402} {\bibfield  {journal} {\bibinfo  {journal}
  {Phys. Rev. Lett.}\ }\textbf {\bibinfo {volume} {116}},\ \bibinfo {pages}
  {030402} (\bibinfo {year} {2016})}\BibitemShut {NoStop}%
\bibitem [{\citenamefont {Tomita}\ \emph {et~al.}(2017)\citenamefont {Tomita},
  \citenamefont {Nakajima}, \citenamefont {Danshita}, \citenamefont {Takasu},\
  and\ \citenamefont {Takahashi}}]{Tomita_2017}%
  \BibitemOpen
  \bibfield  {author} {\bibinfo {author} {\bibfnamefont {T.}~\bibnamefont
  {Tomita}}, \bibinfo {author} {\bibfnamefont {S.}~\bibnamefont {Nakajima}},
  \bibinfo {author} {\bibfnamefont {I.}~\bibnamefont {Danshita}}, \bibinfo
  {author} {\bibfnamefont {Y.}~\bibnamefont {Takasu}}, \ and\ \bibinfo {author}
  {\bibfnamefont {Y.}~\bibnamefont {Takahashi}},\ }\href {\doibase
  10.1126/sciadv.1701513} {\bibfield  {journal} {\bibinfo  {journal} {Sci.
  Adv.}\ }\textbf {\bibinfo {volume} {3}},\ \bibinfo {pages} {e1701513}
  (\bibinfo {year} {2017})}\BibitemShut {NoStop}%
\bibitem [{\citenamefont {Bouganne}\ \emph {et~al.}(2019)\citenamefont
  {Bouganne}, \citenamefont {Bosch~Aguilera}, \citenamefont {Ghermanoui},
  \citenamefont {Beugnon},\ and\ \citenamefont {Gerbier}}]{Bouganne_2019}%
  \BibitemOpen
  \bibfield  {author} {\bibinfo {author} {\bibfnamefont {R.}~\bibnamefont
  {Bouganne}}, \bibinfo {author} {\bibfnamefont {M.}~\bibnamefont
  {Bosch~Aguilera}}, \bibinfo {author} {\bibfnamefont {A.}~\bibnamefont
  {Ghermanoui}}, \bibinfo {author} {\bibfnamefont {J.}~\bibnamefont {Beugnon}},
  \ and\ \bibinfo {author} {\bibfnamefont {F.}~\bibnamefont {Gerbier}},\ }\href
  {\doibase 10.1038/s41567-019-0678-2} {\bibfield  {journal} {\bibinfo
  {journal} {Nat. Phys.}\ }\textbf {\bibinfo {volume} {16}},\ \bibinfo {pages}
  {21} (\bibinfo {year} {2019})}\BibitemShut {NoStop}%
\bibitem [{\citenamefont {Bouchoule}\ and\ \citenamefont
  {Schemmer}(2020)}]{Bouchoule_2020}%
  \BibitemOpen
  \bibfield  {author} {\bibinfo {author} {\bibfnamefont {I.}~\bibnamefont
  {Bouchoule}}\ and\ \bibinfo {author} {\bibfnamefont {M.}~\bibnamefont
  {Schemmer}},\ }\href {\doibase 10.21468/SciPostPhys.8.4.060} {\bibfield
  {journal} {\bibinfo  {journal} {SciPost Phys.}\ }\textbf {\bibinfo {volume}
  {8}},\ \bibinfo {pages} {60} (\bibinfo {year} {2020})}\BibitemShut {NoStop}%
\bibitem [{\citenamefont {Verstraete}\ \emph {et~al.}(2009)\citenamefont
  {Verstraete}, \citenamefont {Wolf},\ and\ \citenamefont
  {Cirac}}]{Verstraete_2009}%
  \BibitemOpen
  \bibfield  {author} {\bibinfo {author} {\bibfnamefont {F.}~\bibnamefont
  {Verstraete}}, \bibinfo {author} {\bibfnamefont {M.~M.}\ \bibnamefont
  {Wolf}}, \ and\ \bibinfo {author} {\bibfnamefont {J.~I.}\ \bibnamefont
  {Cirac}},\ }\href {\doibase 10.1038/nphys1342} {\bibfield  {journal}
  {\bibinfo  {journal} {Nat. Phys.}\ }\textbf {\bibinfo {volume} {5}},\
  \bibinfo {pages} {633} (\bibinfo {year} {2009})}\BibitemShut {NoStop}%
\bibitem [{\citenamefont {Diehl}\ \emph {et~al.}(2008)\citenamefont {Diehl},
  \citenamefont {Micheli}, \citenamefont {Kantian}, \citenamefont {Kraus},
  \citenamefont {B\"uchler},\ and\ \citenamefont {Zoller}}]{Diehl_2008}%
  \BibitemOpen
  \bibfield  {author} {\bibinfo {author} {\bibfnamefont {S.}~\bibnamefont
  {Diehl}}, \bibinfo {author} {\bibfnamefont {A.}~\bibnamefont {Micheli}},
  \bibinfo {author} {\bibfnamefont {A.}~\bibnamefont {Kantian}}, \bibinfo
  {author} {\bibfnamefont {B.}~\bibnamefont {Kraus}}, \bibinfo {author}
  {\bibfnamefont {H.-P.}\ \bibnamefont {B\"uchler}}, \ and\ \bibinfo {author}
  {\bibfnamefont {P.}~\bibnamefont {Zoller}},\ }\href {\doibase
  10.1038/nphys1073} {\bibfield  {journal} {\bibinfo  {journal} {Nat. Phys.}\
  }\textbf {\bibinfo {volume} {4}},\ \bibinfo {pages} {878} (\bibinfo {year}
  {2008})}\BibitemShut {NoStop}%
\bibitem [{\citenamefont {Roncaglia}\ \emph {et~al.}(2010)\citenamefont
  {Roncaglia}, \citenamefont {Rizzi},\ and\ \citenamefont
  {Cirac}}]{Roncaglia_2010}%
  \BibitemOpen
  \bibfield  {author} {\bibinfo {author} {\bibfnamefont {M.}~\bibnamefont
  {Roncaglia}}, \bibinfo {author} {\bibfnamefont {M.}~\bibnamefont {Rizzi}}, \
  and\ \bibinfo {author} {\bibfnamefont {J.~I.}\ \bibnamefont {Cirac}},\ }\href
  {\doibase 10.1103/PhysRevLett.104.096803} {\bibfield  {journal} {\bibinfo
  {journal} {Phys. Rev. Lett.}\ }\textbf {\bibinfo {volume} {104}},\ \bibinfo
  {pages} {096803} (\bibinfo {year} {2010})}\BibitemShut {NoStop}%
\bibitem [{\citenamefont {Gong}\ \emph {et~al.}(2017)\citenamefont {Gong},
  \citenamefont {Higashikawa},\ and\ \citenamefont {Ueda}}]{Gong_2017}%
  \BibitemOpen
  \bibfield  {author} {\bibinfo {author} {\bibfnamefont {Z.}~\bibnamefont
  {Gong}}, \bibinfo {author} {\bibfnamefont {S.}~\bibnamefont {Higashikawa}}, \
  and\ \bibinfo {author} {\bibfnamefont {M.}~\bibnamefont {Ueda}},\ }\href
  {\doibase 10.1103/PhysRevLett.118.200401} {\bibfield  {journal} {\bibinfo
  {journal} {Phys. Rev. Lett.}\ }\textbf {\bibinfo {volume} {118}},\ \bibinfo
  {pages} {200401} (\bibinfo {year} {2017})}\BibitemShut {NoStop}%
\bibitem [{\citenamefont {Schemmer}\ and\ \citenamefont
  {Bouchoule}(2018)}]{Schemmer_2018}%
  \BibitemOpen
  \bibfield  {author} {\bibinfo {author} {\bibfnamefont {M.}~\bibnamefont
  {Schemmer}}\ and\ \bibinfo {author} {\bibfnamefont {I.}~\bibnamefont
  {Bouchoule}},\ }\href {\doibase 10.1103/PhysRevLett.121.200401} {\bibfield
  {journal} {\bibinfo  {journal} {Phys. Rev. Lett.}\ }\textbf {\bibinfo
  {volume} {121}},\ \bibinfo {pages} {200401} (\bibinfo {year}
  {2018})}\BibitemShut {NoStop}%
\bibitem [{\citenamefont {Dogra}\ \emph {et~al.}(2019)\citenamefont {Dogra},
  \citenamefont {Glidden}, \citenamefont {Hilker}, \citenamefont {Eigen},
  \citenamefont {Cornell}, \citenamefont {Smith},\ and\ \citenamefont
  {Hadzibabic}}]{Dogra_2019}%
  \BibitemOpen
  \bibfield  {author} {\bibinfo {author} {\bibfnamefont {L.~H.}\ \bibnamefont
  {Dogra}}, \bibinfo {author} {\bibfnamefont {J.~A.~P.}\ \bibnamefont
  {Glidden}}, \bibinfo {author} {\bibfnamefont {T.~A.}\ \bibnamefont {Hilker}},
  \bibinfo {author} {\bibfnamefont {C.}~\bibnamefont {Eigen}}, \bibinfo
  {author} {\bibfnamefont {E.~A.}\ \bibnamefont {Cornell}}, \bibinfo {author}
  {\bibfnamefont {R.~P.}\ \bibnamefont {Smith}}, \ and\ \bibinfo {author}
  {\bibfnamefont {Z.}~\bibnamefont {Hadzibabic}},\ }\href {\doibase
  10.1103/PhysRevLett.123.020405} {\bibfield  {journal} {\bibinfo  {journal}
  {Phys. Rev. Lett.}\ }\textbf {\bibinfo {volume} {123}},\ \bibinfo {pages}
  {020405} (\bibinfo {year} {2019})}\BibitemShut {NoStop}%
\bibitem [{\citenamefont {Ashida}\ \emph {et~al.}(2020)\citenamefont {Ashida},
  \citenamefont {Gong},\ and\ \citenamefont {Ueda}}]{Nakagawa_2020}%
  \BibitemOpen
  \bibfield  {author} {\bibinfo {author} {\bibfnamefont {Y.}~\bibnamefont
  {Ashida}}, \bibinfo {author} {\bibfnamefont {Z.}~\bibnamefont {Gong}}, \ and\
  \bibinfo {author} {\bibfnamefont {M.}~\bibnamefont {Ueda}},\ }\href@noop {}
  {\enquote {\bibinfo {title} {Non-hermitian physics},}\ } (\bibinfo {year}
  {2020}),\ \Eprint {http://arxiv.org/abs/2003.14202} {arXiv:2003.14202
  [cond-mat.quant-gas]} \BibitemShut {NoStop}%
\bibitem [{\citenamefont {Daley}\ \emph {et~al.}(2009)\citenamefont {Daley},
  \citenamefont {Taylor}, \citenamefont {Diehl}, \citenamefont {Baranov},\ and\
  \citenamefont {Zoller}}]{Daley_2009}%
  \BibitemOpen
  \bibfield  {author} {\bibinfo {author} {\bibfnamefont {A.~J.}\ \bibnamefont
  {Daley}}, \bibinfo {author} {\bibfnamefont {J.~M.}\ \bibnamefont {Taylor}},
  \bibinfo {author} {\bibfnamefont {S.}~\bibnamefont {Diehl}}, \bibinfo
  {author} {\bibfnamefont {M.}~\bibnamefont {Baranov}}, \ and\ \bibinfo
  {author} {\bibfnamefont {P.}~\bibnamefont {Zoller}},\ }\href {\doibase
  10.1103/PhysRevLett.102.040402} {\bibfield  {journal} {\bibinfo  {journal}
  {Phys. Rev. Lett.}\ }\textbf {\bibinfo {volume} {102}},\ \bibinfo {pages}
  {040402} (\bibinfo {year} {2009})}\BibitemShut {NoStop}%
\bibitem [{\citenamefont {Kantian}\ \emph {et~al.}(2009)\citenamefont
  {Kantian}, \citenamefont {Dalmonte}, \citenamefont {Diehl}, \citenamefont
  {Hofstetter}, \citenamefont {Zoller},\ and\ \citenamefont
  {Daley}}]{Kantian_2009}%
  \BibitemOpen
  \bibfield  {author} {\bibinfo {author} {\bibfnamefont {A.}~\bibnamefont
  {Kantian}}, \bibinfo {author} {\bibfnamefont {M.}~\bibnamefont {Dalmonte}},
  \bibinfo {author} {\bibfnamefont {S.}~\bibnamefont {Diehl}}, \bibinfo
  {author} {\bibfnamefont {W.}~\bibnamefont {Hofstetter}}, \bibinfo {author}
  {\bibfnamefont {P.}~\bibnamefont {Zoller}}, \ and\ \bibinfo {author}
  {\bibfnamefont {A.~J.}\ \bibnamefont {Daley}},\ }\href {\doibase
  10.1103/PhysRevLett.103.240401} {\bibfield  {journal} {\bibinfo  {journal}
  {Phys. Rev. Lett.}\ }\textbf {\bibinfo {volume} {103}},\ \bibinfo {pages}
  {240401} (\bibinfo {year} {2009})}\BibitemShut {NoStop}%
\bibitem [{\citenamefont {Foss-Feig}\ \emph {et~al.}(2012)\citenamefont
  {Foss-Feig}, \citenamefont {Daley}, \citenamefont {Thompson},\ and\
  \citenamefont {Rey}}]{FossFeig_2012}%
  \BibitemOpen
  \bibfield  {author} {\bibinfo {author} {\bibfnamefont {M.}~\bibnamefont
  {Foss-Feig}}, \bibinfo {author} {\bibfnamefont {A.~J.}\ \bibnamefont
  {Daley}}, \bibinfo {author} {\bibfnamefont {J.~K.}\ \bibnamefont {Thompson}},
  \ and\ \bibinfo {author} {\bibfnamefont {A.~M.}\ \bibnamefont {Rey}},\ }\href
  {\doibase 10.1103/PhysRevLett.109.230501} {\bibfield  {journal} {\bibinfo
  {journal} {Phys. Rev. Lett.}\ }\textbf {\bibinfo {volume} {109}},\ \bibinfo
  {pages} {230501} (\bibinfo {year} {2012})}\BibitemShut {NoStop}%
\bibitem [{\citenamefont {Syassen}\ \emph {et~al.}(2008)\citenamefont
  {Syassen}, \citenamefont {Bauer}, \citenamefont {Lettner}, \citenamefont
  {Volz}, \citenamefont {Dietze}, \citenamefont {Garc{\'\i}a-Ripoll},
  \citenamefont {Cirac}, \citenamefont {Rempe},\ and\ \citenamefont
  {D{\"u}rr}}]{Syassen_2008}%
  \BibitemOpen
  \bibfield  {author} {\bibinfo {author} {\bibfnamefont {N.}~\bibnamefont
  {Syassen}}, \bibinfo {author} {\bibfnamefont {D.~M.}\ \bibnamefont {Bauer}},
  \bibinfo {author} {\bibfnamefont {M.}~\bibnamefont {Lettner}}, \bibinfo
  {author} {\bibfnamefont {T.}~\bibnamefont {Volz}}, \bibinfo {author}
  {\bibfnamefont {D.}~\bibnamefont {Dietze}}, \bibinfo {author} {\bibfnamefont
  {J.~J.}\ \bibnamefont {Garc{\'\i}a-Ripoll}}, \bibinfo {author} {\bibfnamefont
  {J.~I.}\ \bibnamefont {Cirac}}, \bibinfo {author} {\bibfnamefont
  {G.}~\bibnamefont {Rempe}}, \ and\ \bibinfo {author} {\bibfnamefont
  {S.}~\bibnamefont {D{\"u}rr}},\ }\href {\doibase 10.1126/science.1155309}
  {\bibfield  {journal} {\bibinfo  {journal} {Science}\ }\textbf {\bibinfo
  {volume} {320}},\ \bibinfo {pages} {1329} (\bibinfo {year}
  {2008})}\BibitemShut {NoStop}%
\bibitem [{\citenamefont {Garc{\'{\i}}a-Ripoll}\ \emph
  {et~al.}(2009)\citenamefont {Garc{\'{\i}}a-Ripoll}, \citenamefont {D\"urr},
  \citenamefont {Syassen}, \citenamefont {Bauer}, \citenamefont {Lettner},
  \citenamefont {Rempe},\ and\ \citenamefont {Cirac}}]{GarciaRipoll_2009}%
  \BibitemOpen
  \bibfield  {author} {\bibinfo {author} {\bibfnamefont {J.~J.}\ \bibnamefont
  {Garc{\'{\i}}a-Ripoll}}, \bibinfo {author} {\bibfnamefont {S.}~\bibnamefont
  {D\"urr}}, \bibinfo {author} {\bibfnamefont {N.}~\bibnamefont {Syassen}},
  \bibinfo {author} {\bibfnamefont {D.~M.}\ \bibnamefont {Bauer}}, \bibinfo
  {author} {\bibfnamefont {M.}~\bibnamefont {Lettner}}, \bibinfo {author}
  {\bibfnamefont {G.}~\bibnamefont {Rempe}}, \ and\ \bibinfo {author}
  {\bibfnamefont {J.~I.}\ \bibnamefont {Cirac}},\ }\href {\doibase
  10.1088/1367-2630/11/1/013053} {\bibfield  {journal} {\bibinfo  {journal}
  {New J. Phys.}\ }\textbf {\bibinfo {volume} {11}},\ \bibinfo {pages} {013053}
  (\bibinfo {year} {2009})}\BibitemShut {NoStop}%
\bibitem [{\citenamefont {Tomita}\ \emph {et~al.}(2019)\citenamefont {Tomita},
  \citenamefont {Nakajima}, \citenamefont {Takasu},\ and\ \citenamefont
  {Takahashi}}]{Tomita_2019}%
  \BibitemOpen
  \bibfield  {author} {\bibinfo {author} {\bibfnamefont {T.}~\bibnamefont
  {Tomita}}, \bibinfo {author} {\bibfnamefont {S.}~\bibnamefont {Nakajima}},
  \bibinfo {author} {\bibfnamefont {Y.}~\bibnamefont {Takasu}}, \ and\ \bibinfo
  {author} {\bibfnamefont {Y.}~\bibnamefont {Takahashi}},\ }\href {\doibase
  10.1103/PhysRevA.99.031601} {\bibfield  {journal} {\bibinfo  {journal} {Phys.
  Rev. A}\ }\textbf {\bibinfo {volume} {99}},\ \bibinfo {pages} {031601}
  (\bibinfo {year} {2019})}\BibitemShut {NoStop}%
\bibitem [{\citenamefont {Zhu}\ \emph {et~al.}(2014)\citenamefont {Zhu},
  \citenamefont {Gadway}, \citenamefont {Foss-Feig}, \citenamefont
  {Schachenmayer}, \citenamefont {Wall}, \citenamefont {Hazzard}, \citenamefont
  {Yan}, \citenamefont {Moses}, \citenamefont {Covey}, \citenamefont {Jin},
  \citenamefont {Ye}, \citenamefont {Holland},\ and\ \citenamefont
  {Rey}}]{Zhu_2014}%
  \BibitemOpen
  \bibfield  {author} {\bibinfo {author} {\bibfnamefont {B.}~\bibnamefont
  {Zhu}}, \bibinfo {author} {\bibfnamefont {B.}~\bibnamefont {Gadway}},
  \bibinfo {author} {\bibfnamefont {M.}~\bibnamefont {Foss-Feig}}, \bibinfo
  {author} {\bibfnamefont {J.}~\bibnamefont {Schachenmayer}}, \bibinfo {author}
  {\bibfnamefont {M.~L.}\ \bibnamefont {Wall}}, \bibinfo {author}
  {\bibfnamefont {K.~R.~A.}\ \bibnamefont {Hazzard}}, \bibinfo {author}
  {\bibfnamefont {B.}~\bibnamefont {Yan}}, \bibinfo {author} {\bibfnamefont
  {S.~A.}\ \bibnamefont {Moses}}, \bibinfo {author} {\bibfnamefont {J.~P.}\
  \bibnamefont {Covey}}, \bibinfo {author} {\bibfnamefont {D.~S.}\ \bibnamefont
  {Jin}}, \bibinfo {author} {\bibfnamefont {J.}~\bibnamefont {Ye}}, \bibinfo
  {author} {\bibfnamefont {M.}~\bibnamefont {Holland}}, \ and\ \bibinfo
  {author} {\bibfnamefont {A.~M.}\ \bibnamefont {Rey}},\ }\href {\doibase
  10.1103/PhysRevLett.112.070404} {\bibfield  {journal} {\bibinfo  {journal}
  {Phys. Rev. Lett.}\ }\textbf {\bibinfo {volume} {112}},\ \bibinfo {pages}
  {070404} (\bibinfo {year} {2014})}\BibitemShut {NoStop}%
\bibitem [{\citenamefont {Sponselee}\ \emph {et~al.}(2019)\citenamefont
  {Sponselee}, \citenamefont {Freystatzky}, \citenamefont {Abeln},
  \citenamefont {Diem}, \citenamefont {Hundt}, \citenamefont {Kochanke},
  \citenamefont {Ponath}, \citenamefont {Santra}, \citenamefont {Mathey},
  \citenamefont {Sengstock},\ and\ \citenamefont {Becker}}]{Sponselee_2018}%
  \BibitemOpen
  \bibfield  {author} {\bibinfo {author} {\bibfnamefont {K.}~\bibnamefont
  {Sponselee}}, \bibinfo {author} {\bibfnamefont {L.}~\bibnamefont
  {Freystatzky}}, \bibinfo {author} {\bibfnamefont {B.}~\bibnamefont {Abeln}},
  \bibinfo {author} {\bibfnamefont {M.}~\bibnamefont {Diem}}, \bibinfo {author}
  {\bibfnamefont {B.}~\bibnamefont {Hundt}}, \bibinfo {author} {\bibfnamefont
  {A.}~\bibnamefont {Kochanke}}, \bibinfo {author} {\bibfnamefont
  {T.}~\bibnamefont {Ponath}}, \bibinfo {author} {\bibfnamefont
  {B.}~\bibnamefont {Santra}}, \bibinfo {author} {\bibfnamefont
  {L.}~\bibnamefont {Mathey}}, \bibinfo {author} {\bibfnamefont
  {K.}~\bibnamefont {Sengstock}}, \ and\ \bibinfo {author} {\bibfnamefont
  {C.}~\bibnamefont {Becker}},\ }\href {\doibase 10.1088/2058-9565/aadccd}
  {\bibfield  {journal} {\bibinfo  {journal} {Quantum Sci. Technol.}\ }\textbf
  {\bibinfo {volume} {4}},\ \bibinfo {pages} {014002} (\bibinfo {year}
  {2019})}\BibitemShut {NoStop}%
\bibitem [{\citenamefont {Mark}\ \emph {et~al.}(2020)\citenamefont {Mark},
  \citenamefont {Flannigan}, \citenamefont {Meinert}, \citenamefont {D'Incao},
  \citenamefont {Daley},\ and\ \citenamefont {N\"agerl}}]{Mark_2020}%
  \BibitemOpen
  \bibfield  {author} {\bibinfo {author} {\bibfnamefont {M.~J.}\ \bibnamefont
  {Mark}}, \bibinfo {author} {\bibfnamefont {S.}~\bibnamefont {Flannigan}},
  \bibinfo {author} {\bibfnamefont {F.}~\bibnamefont {Meinert}}, \bibinfo
  {author} {\bibfnamefont {J.~P.}\ \bibnamefont {D'Incao}}, \bibinfo {author}
  {\bibfnamefont {A.~J.}\ \bibnamefont {Daley}}, \ and\ \bibinfo {author}
  {\bibfnamefont {H.-C.}\ \bibnamefont {N\"agerl}},\ }\href {\doibase
  10.1103/PhysRevResearch.2.043050} {\bibfield  {journal} {\bibinfo  {journal}
  {Phys. Rev. Research}\ }\textbf {\bibinfo {volume} {2}},\ \bibinfo {pages}
  {043050} (\bibinfo {year} {2020})}\BibitemShut {NoStop}%
\bibitem [{\citenamefont {Langen}\ \emph {et~al.}(2015)\citenamefont {Langen},
  \citenamefont {Erne}, \citenamefont {Geiger}, \citenamefont {Rauer},
  \citenamefont {Schweigler}, \citenamefont {Kuhnert}, \citenamefont
  {Rohringer}, \citenamefont {Mazets}, \citenamefont {Gasenzer},\ and\
  \citenamefont {Schmiedmayer}}]{Langen_2015}%
  \BibitemOpen
  \bibfield  {author} {\bibinfo {author} {\bibfnamefont {T.}~\bibnamefont
  {Langen}}, \bibinfo {author} {\bibfnamefont {S.}~\bibnamefont {Erne}},
  \bibinfo {author} {\bibfnamefont {R.}~\bibnamefont {Geiger}}, \bibinfo
  {author} {\bibfnamefont {B.}~\bibnamefont {Rauer}}, \bibinfo {author}
  {\bibfnamefont {T.}~\bibnamefont {Schweigler}}, \bibinfo {author}
  {\bibfnamefont {M.}~\bibnamefont {Kuhnert}}, \bibinfo {author} {\bibfnamefont
  {W.}~\bibnamefont {Rohringer}}, \bibinfo {author} {\bibfnamefont {I.~E.}\
  \bibnamefont {Mazets}}, \bibinfo {author} {\bibfnamefont {T.}~\bibnamefont
  {Gasenzer}}, \ and\ \bibinfo {author} {\bibfnamefont {J.}~\bibnamefont
  {Schmiedmayer}},\ }\href {\doibase 10.1126/science.1257026} {\bibfield
  {journal} {\bibinfo  {journal} {Science}\ }\textbf {\bibinfo {volume}
  {348}},\ \bibinfo {pages} {207} (\bibinfo {year} {2015})}\BibitemShut
  {NoStop}%
\bibitem [{\citenamefont {Essler}\ and\ \citenamefont
  {Fagotti}(2016)}]{Essler_2016}%
  \BibitemOpen
  \bibfield  {author} {\bibinfo {author} {\bibfnamefont {F.~H.~L.}\
  \bibnamefont {Essler}}\ and\ \bibinfo {author} {\bibfnamefont
  {M.}~\bibnamefont {Fagotti}},\ }\href {\doibase
  10.1088/1742-5468/2016/06/064002} {\bibfield  {journal} {\bibinfo  {journal}
  {J. Stat. Mech.}\ ,\ \bibinfo {pages} {064002}} (\bibinfo {year}
  {2016})}\BibitemShut {NoStop}%
\bibitem [{\citenamefont {Cazalilla}\ and\ \citenamefont
  {Chung}(2016)}]{Cazalilla_2016}%
  \BibitemOpen
  \bibfield  {author} {\bibinfo {author} {\bibfnamefont {M.~A.}\ \bibnamefont
  {Cazalilla}}\ and\ \bibinfo {author} {\bibfnamefont {M.-C.}\ \bibnamefont
  {Chung}},\ }\href {\doibase 10.1088/1742-5468/2016/06/064004} {\bibfield
  {journal} {\bibinfo  {journal} {J. Stat. Mech.}\ ,\ \bibinfo {pages}
  {064004}} (\bibinfo {year} {2016})}\BibitemShut {NoStop}%
\bibitem [{\citenamefont {Vidmar}\ and\ \citenamefont
  {Rigol}(2016)}]{Vidmar_2016}%
  \BibitemOpen
  \bibfield  {author} {\bibinfo {author} {\bibfnamefont {L.}~\bibnamefont
  {Vidmar}}\ and\ \bibinfo {author} {\bibfnamefont {M.}~\bibnamefont {Rigol}},\
  }\href {\doibase 10.1088/1742-5468/2016/06/064007} {\bibfield  {journal}
  {\bibinfo  {journal} {J. Stat. Mech.}\ ,\ \bibinfo {pages} {064007}}
  (\bibinfo {year} {2016})}\BibitemShut {NoStop}%
\bibitem [{\citenamefont {Langen}\ \emph {et~al.}(2016)\citenamefont {Langen},
  \citenamefont {Gasenzer},\ and\ \citenamefont {Schmiedmayer}}]{Langen_2016}%
  \BibitemOpen
  \bibfield  {author} {\bibinfo {author} {\bibfnamefont {T.}~\bibnamefont
  {Langen}}, \bibinfo {author} {\bibfnamefont {T.}~\bibnamefont {Gasenzer}}, \
  and\ \bibinfo {author} {\bibfnamefont {J.}~\bibnamefont {Schmiedmayer}},\
  }\href {\doibase 10.1088/1742-5468/2016/06/064009} {\bibfield  {journal}
  {\bibinfo  {journal} {J. Stat. Mech.}\ ,\ \bibinfo {pages} {064009}}
  (\bibinfo {year} {2016})}\BibitemShut {NoStop}%
\bibitem [{\citenamefont {Lange}\ \emph {et~al.}(2018)\citenamefont {Lange},
  \citenamefont {Lenar\u{c}i\u{c}},\ and\ \citenamefont {Rosch}}]{Lange_2018}%
  \BibitemOpen
  \bibfield  {author} {\bibinfo {author} {\bibfnamefont {F.}~\bibnamefont
  {Lange}}, \bibinfo {author} {\bibfnamefont {Z.}~\bibnamefont
  {Lenar\u{c}i\u{c}}}, \ and\ \bibinfo {author} {\bibfnamefont
  {A.}~\bibnamefont {Rosch}},\ }\href {\doibase 10.1103/PhysRevB.97.165138}
  {\bibfield  {journal} {\bibinfo  {journal} {Phys. Rev. B}\ }\textbf {\bibinfo
  {volume} {97}},\ \bibinfo {pages} {165138} (\bibinfo {year}
  {2018})}\BibitemShut {NoStop}%
\bibitem [{\citenamefont {Mallayya}\ \emph {et~al.}(2019)\citenamefont
  {Mallayya}, \citenamefont {Rigol},\ and\ \citenamefont
  {De~Roeck}}]{Mallaya_2019}%
  \BibitemOpen
  \bibfield  {author} {\bibinfo {author} {\bibfnamefont {K.}~\bibnamefont
  {Mallayya}}, \bibinfo {author} {\bibfnamefont {M.}~\bibnamefont {Rigol}}, \
  and\ \bibinfo {author} {\bibfnamefont {W.}~\bibnamefont {De~Roeck}},\ }\href
  {\doibase 10.1103/PhysRevX.9.021027} {\bibfield  {journal} {\bibinfo
  {journal} {Phys. Rev. X}\ }\textbf {\bibinfo {volume} {9}},\ \bibinfo {pages}
  {021027} (\bibinfo {year} {2019})}\BibitemShut {NoStop}%
\bibitem [{\citenamefont {Caux}\ \emph {et~al.}(2019)\citenamefont {Caux},
  \citenamefont {Doyon}, \citenamefont {Dubail}, \citenamefont {Konik},\ and\
  \citenamefont {Yoshimura}}]{Caux_2019}%
  \BibitemOpen
  \bibfield  {author} {\bibinfo {author} {\bibfnamefont {J.-S.}\ \bibnamefont
  {Caux}}, \bibinfo {author} {\bibfnamefont {B.}~\bibnamefont {Doyon}},
  \bibinfo {author} {\bibfnamefont {J.}~\bibnamefont {Dubail}}, \bibinfo
  {author} {\bibfnamefont {R.}~\bibnamefont {Konik}}, \ and\ \bibinfo {author}
  {\bibfnamefont {T.}~\bibnamefont {Yoshimura}},\ }\href {\doibase
  10.21468/SciPostPhys.6.6.070} {\bibfield  {journal} {\bibinfo  {journal}
  {SciPost Phys.}\ }\textbf {\bibinfo {volume} {6}},\ \bibinfo {pages} {70}
  (\bibinfo {year} {2019})}\BibitemShut {NoStop}%
\bibitem [{\citenamefont {Schemmer}\ \emph {et~al.}(2019)\citenamefont
  {Schemmer}, \citenamefont {Bouchoule}, \citenamefont {Doyon},\ and\
  \citenamefont {Dubail}}]{Schemmer_2019}%
  \BibitemOpen
  \bibfield  {author} {\bibinfo {author} {\bibfnamefont {M.}~\bibnamefont
  {Schemmer}}, \bibinfo {author} {\bibfnamefont {I.}~\bibnamefont {Bouchoule}},
  \bibinfo {author} {\bibfnamefont {B.}~\bibnamefont {Doyon}}, \ and\ \bibinfo
  {author} {\bibfnamefont {J.}~\bibnamefont {Dubail}},\ }\href {\doibase
  10.1103/PhysRevLett.122.090601} {\bibfield  {journal} {\bibinfo  {journal}
  {Phys. Rev. Lett.}\ }\textbf {\bibinfo {volume} {122}},\ \bibinfo {pages}
  {090601} (\bibinfo {year} {2019})}\BibitemShut {NoStop}%
\bibitem [{Note1()}]{Note1}%
  \BibitemOpen
  \bibinfo {note} {For $J=0$, the population of doubly-occupied sites decays as
  $p_2(t) = p_2(0) e^{- \gamma _\protect \mathrm {2B} t}$.}\BibitemShut {Stop}%
\bibitem [{\citenamefont {Daley}(2014)}]{Daley_2014}%
  \BibitemOpen
  \bibfield  {author} {\bibinfo {author} {\bibfnamefont {A.~J.}\ \bibnamefont
  {Daley}},\ }\href {\doibase 10.1080/00018732.2014.933502} {\bibfield
  {journal} {\bibinfo  {journal} {Adv. Phys.}\ }\textbf {\bibinfo {volume}
  {63}},\ \bibinfo {pages} {77} (\bibinfo {year} {2014})}\BibitemShut {NoStop}%
\bibitem [{\citenamefont {Jordan}\ and\ \citenamefont
  {Wigner}(1928)}]{JordanWigner_1928}%
  \BibitemOpen
  \bibfield  {author} {\bibinfo {author} {\bibfnamefont {P.}~\bibnamefont
  {Jordan}}\ and\ \bibinfo {author} {\bibfnamefont {E.~P.}\ \bibnamefont
  {Wigner}},\ }\href@noop {} {\bibfield  {journal} {\bibinfo  {journal} {Z.
  Phys.}\ }\textbf {\bibinfo {volume} {47}},\ \bibinfo {pages} {631} (\bibinfo
  {year} {1928})}\BibitemShut {NoStop}%
\bibitem [{\citenamefont {Sotiriadis}\ and\ \citenamefont
  {Calabrese}(2014)}]{Sotiriadis_2014}%
  \BibitemOpen
  \bibfield  {author} {\bibinfo {author} {\bibfnamefont {S.}~\bibnamefont
  {Sotiriadis}}\ and\ \bibinfo {author} {\bibfnamefont {P.}~\bibnamefont
  {Calabrese}},\ }\href {\doibase 10.1088/1742-5468/2014/07/p07024} {\bibfield
  {journal} {\bibinfo  {journal} {J. Stat. Mech.}\ ,\ \bibinfo {pages}
  {P07024}} (\bibinfo {year} {2014})}\BibitemShut {NoStop}%
\bibitem [{Sup()}]{SupMat}%
  \BibitemOpen
  \href@noop {} {\ }\bibinfo {note} {See Supplementary Material.}\BibitemShut
  {Stop}%
\bibitem [{\citenamefont {Gangardt}\ and\ \citenamefont
  {Shlyapnikov}(2006)}]{Gangardt_2006}%
  \BibitemOpen
  \bibfield  {author} {\bibinfo {author} {\bibfnamefont {D.~M.}\ \bibnamefont
  {Gangardt}}\ and\ \bibinfo {author} {\bibfnamefont {G.~V.}\ \bibnamefont
  {Shlyapnikov}},\ }\href {\doibase 10.1088/1367-2630/8/8/167} {\bibfield
  {journal} {\bibinfo  {journal} {New J. Phys.}\ }\textbf {\bibinfo {volume}
  {8}},\ \bibinfo {pages} {167} (\bibinfo {year} {2006})}\BibitemShut {NoStop}%
\bibitem [{\citenamefont {Lange}\ \emph {et~al.}(2017)\citenamefont {Lange},
  \citenamefont {Lenar\u{c}i\u{c}},\ and\ \citenamefont {Rosch}}]{Lange_2017}%
  \BibitemOpen
  \bibfield  {author} {\bibinfo {author} {\bibfnamefont {F.}~\bibnamefont
  {Lange}}, \bibinfo {author} {\bibfnamefont {Z.}~\bibnamefont
  {Lenar\u{c}i\u{c}}}, \ and\ \bibinfo {author} {\bibfnamefont
  {A.}~\bibnamefont {Rosch}},\ }\href {\doibase 10.1038/ncomms15767} {\bibfield
   {journal} {\bibinfo  {journal} {Nat. Commun.}\ }\textbf {\bibinfo {volume}
  {8}},\ \bibinfo {pages} {15767} (\bibinfo {year} {2017})}\BibitemShut
  {NoStop}%
\bibitem [{\citenamefont {Lenar\u{c}i\u{c}}\ \emph {et~al.}(2018)\citenamefont
  {Lenar\u{c}i\u{c}}, \citenamefont {Lange},\ and\ \citenamefont
  {Rosch}}]{Lenarcic_2018}%
  \BibitemOpen
  \bibfield  {author} {\bibinfo {author} {\bibfnamefont {Z.}~\bibnamefont
  {Lenar\u{c}i\u{c}}}, \bibinfo {author} {\bibfnamefont {F.}~\bibnamefont
  {Lange}}, \ and\ \bibinfo {author} {\bibfnamefont {A.}~\bibnamefont
  {Rosch}},\ }\href {\doibase 10.1103/PhysRevB.97.024302} {\bibfield  {journal}
  {\bibinfo  {journal} {Phys. Rev. B}\ }\textbf {\bibinfo {volume} {97}},\
  \bibinfo {pages} {024302} (\bibinfo {year} {2018})}\BibitemShut {NoStop}%
\bibitem [{\citenamefont {Girardeau}(1960)}]{Girardeau_1960}%
  \BibitemOpen
  \bibfield  {author} {\bibinfo {author} {\bibfnamefont {M.}~\bibnamefont
  {Girardeau}},\ }\href {\doibase 10.1063/1.1703687} {\bibfield  {journal}
  {\bibinfo  {journal} {J. Math. Phys.}\ }\textbf {\bibinfo {volume} {1}},\
  \bibinfo {pages} {516} (\bibinfo {year} {1960})}\BibitemShut {NoStop}%
\bibitem [{\citenamefont {Bouganne}\ \emph {et~al.}(2017)\citenamefont
  {Bouganne}, \citenamefont {Aguilera}, \citenamefont {Dareau}, \citenamefont
  {Soave}, \citenamefont {Beugnon},\ and\ \citenamefont
  {Gerbier}}]{Bouganne_2017}%
  \BibitemOpen
  \bibfield  {author} {\bibinfo {author} {\bibfnamefont {R.}~\bibnamefont
  {Bouganne}}, \bibinfo {author} {\bibfnamefont {M.~B.}\ \bibnamefont
  {Aguilera}}, \bibinfo {author} {\bibfnamefont {A.}~\bibnamefont {Dareau}},
  \bibinfo {author} {\bibfnamefont {E.}~\bibnamefont {Soave}}, \bibinfo
  {author} {\bibfnamefont {J.}~\bibnamefont {Beugnon}}, \ and\ \bibinfo
  {author} {\bibfnamefont {F.}~\bibnamefont {Gerbier}},\ }\href {\doibase
  10.1088/1367-2630/aa8c45} {\bibfield  {journal} {\bibinfo  {journal} {New J.
  Phys.}\ }\textbf {\bibinfo {volume} {19}},\ \bibinfo {pages} {113006}
  (\bibinfo {year} {2017})}\BibitemShut {NoStop}%
\bibitem [{\citenamefont {Franchi}\ \emph {et~al.}(2017)\citenamefont
  {Franchi}, \citenamefont {Livi}, \citenamefont {Cappellini}, \citenamefont
  {Binella}, \citenamefont {Inguscio}, \citenamefont {Catani},\ and\
  \citenamefont {Fallani}}]{Franchi_2017}%
  \BibitemOpen
  \bibfield  {author} {\bibinfo {author} {\bibfnamefont {L.}~\bibnamefont
  {Franchi}}, \bibinfo {author} {\bibfnamefont {L.~F.}\ \bibnamefont {Livi}},
  \bibinfo {author} {\bibfnamefont {G.}~\bibnamefont {Cappellini}}, \bibinfo
  {author} {\bibfnamefont {G.}~\bibnamefont {Binella}}, \bibinfo {author}
  {\bibfnamefont {M.}~\bibnamefont {Inguscio}}, \bibinfo {author}
  {\bibfnamefont {J.}~\bibnamefont {Catani}}, \ and\ \bibinfo {author}
  {\bibfnamefont {L.}~\bibnamefont {Fallani}},\ }\href {\doibase
  10.1088/1367-2630/aa8fb4} {\bibfield  {journal} {\bibinfo  {journal} {New J.
  Phys.}\ }\textbf {\bibinfo {volume} {19}},\ \bibinfo {pages} {103037}
  (\bibinfo {year} {2017})}\BibitemShut {NoStop}%
\bibitem [{\citenamefont {Mazurenko}\ \emph {et~al.}(2017)\citenamefont
  {Mazurenko}, \citenamefont {Chiu}, \citenamefont {Ji}, \citenamefont
  {Parsons}, \citenamefont {Kanász-Nagy}, \citenamefont {Schmidt},
  \citenamefont {Grusdt}, \citenamefont {Demler}, \citenamefont {Greif},\ and\
  \citenamefont {Greiner}}]{Mazurenko2017a}%
  \BibitemOpen
  \bibfield  {author} {\bibinfo {author} {\bibfnamefont {A.}~\bibnamefont
  {Mazurenko}}, \bibinfo {author} {\bibfnamefont {C.~S.}\ \bibnamefont {Chiu}},
  \bibinfo {author} {\bibfnamefont {G.}~\bibnamefont {Ji}}, \bibinfo {author}
  {\bibfnamefont {M.~F.}\ \bibnamefont {Parsons}}, \bibinfo {author}
  {\bibfnamefont {M.}~\bibnamefont {Kanász-Nagy}}, \bibinfo {author}
  {\bibfnamefont {R.}~\bibnamefont {Schmidt}}, \bibinfo {author} {\bibfnamefont
  {F.}~\bibnamefont {Grusdt}}, \bibinfo {author} {\bibfnamefont
  {E.}~\bibnamefont {Demler}}, \bibinfo {author} {\bibfnamefont
  {D.}~\bibnamefont {Greif}}, \ and\ \bibinfo {author} {\bibfnamefont
  {M.}~\bibnamefont {Greiner}},\ }\href {https://doi.org/10.1038/nature22362}
  {\bibfield  {journal} {\bibinfo  {journal} {Nature}\ }\textbf {\bibinfo
  {volume} {545}},\ \bibinfo {pages} {462} (\bibinfo {year}
  {2017})}\BibitemShut {NoStop}%
\bibitem [{\citenamefont {Bloch}\ \emph {et~al.}(2008)\citenamefont {Bloch},
  \citenamefont {Dalibard},\ and\ \citenamefont {Zwerger}}]{Bloch2008a}%
  \BibitemOpen
  \bibfield  {author} {\bibinfo {author} {\bibfnamefont {I.}~\bibnamefont
  {Bloch}}, \bibinfo {author} {\bibfnamefont {J.}~\bibnamefont {Dalibard}}, \
  and\ \bibinfo {author} {\bibfnamefont {W.}~\bibnamefont {Zwerger}},\ }\href
  {\doibase 10.1103/RevModPhys.80.885} {\bibfield  {journal} {\bibinfo
  {journal} {Rev. Mod. Phys.}\ }\textbf {\bibinfo {volume} {80}},\ \bibinfo
  {pages} {885} (\bibinfo {year} {2008})}\BibitemShut {NoStop}%
\bibitem [{\citenamefont {Bouchoule}\ \emph {et~al.}(2020)\citenamefont
  {Bouchoule}, \citenamefont {Doyon},\ and\ \citenamefont
  {Dubail}}]{Bouchoule_2020b}%
  \BibitemOpen
  \bibfield  {author} {\bibinfo {author} {\bibfnamefont {I.}~\bibnamefont
  {Bouchoule}}, \bibinfo {author} {\bibfnamefont {B.}~\bibnamefont {Doyon}}, \
  and\ \bibinfo {author} {\bibfnamefont {J.}~\bibnamefont {Dubail}},\ }\href
  {\doibase 10.21468/SciPostPhys.9.4.044} {\bibfield  {journal} {\bibinfo
  {journal} {SciPost Phys.}\ }\textbf {\bibinfo {volume} {9}},\ \bibinfo
  {pages} {44} (\bibinfo {year} {2020})}\BibitemShut {NoStop}%
\end{thebibliography}
\end{document}